\documentclass[%
 aip,
 amsmath,amssymb,
 reprint,%
]{revtex4-1}
\usepackage[english]{babel}
\usepackage{amsmath}
\usepackage{amssymb}
\usepackage{mathrsfs}
\usepackage{color}
\usepackage{graphicx}
\usepackage{tabularx}
\usepackage{dcolumn} % Align table columns on decimal point
\usepackage{bm}      % bold math
\usepackage{array}   % table making
\usepackage{amsfonts}
\usepackage{bm}
\usepackage{amsbsy}
\usepackage{url}
\usepackage{xr-hyper}
\usepackage{subfig}
\usepackage[bookmarks=false,colorlinks,citecolor=magenta]{hyperref}

\usepackage[utf8]{inputenc}
\usepackage[T1]{fontenc}
\usepackage{mathptmx}

\usepackage{physics}
\usepackage[version=4]{mhchem}
\usepackage{xfrac}
\usepackage{booktabs} % nice table rule
\usepackage{multirow} % multirow in tables

\include{MyCommand}

\DeclareUnicodeCharacter{2212}{-}

\makeatletter
\newcommand*{\addFileDependency}[1]{
  \typeout{(#1)}
  \@addtofilelist{#1}
  \IfFileExists{#1}{}{\typeout{No file #1.}}
}
\makeatother

\begin{document}
\title{Modeling Stochastic Chemical Kinetics on Quantum Computers}

\author{Tilas Kabengele}
\affiliation{Department of Chemistry, Brown University, Providence, Rhode Island 02912, USA}
\affiliation{School of Engineering, Brown University, Providence, Rhode Island 02912, USA}
\author{Yash M. Lokare}
\affiliation{Department of Physics, Brown University, Providence, Rhode Island 02912, USA}
\author{J. B. Marston}
\affiliation{Department of Physics, Brown University, Providence, Rhode Island 02912, USA}
\affiliation{Brown Theoretical Physics Center, Brown University, Providence, Rhode Island 02912, USA}
\author{Brenda M. Rubenstein}
\email{Author to whom correspondence should be addressed: Brenda Rubenstein, brenda\_rubenstein@brown.edu}
\affiliation{Department of Chemistry, Brown University, Providence, Rhode Island 02912, USA}
\affiliation{Department of Physics, Brown University, Providence, Rhode Island 02912, USA}
 
\date{\today}

\begin{abstract}
The Chemical Master Equation (CME) provides a highly accurate, yet extremely resource-intensive representation of a stochastic chemical reaction network and its kinetics due to the exponential scaling of its possible states with the number of reacting species. In this work, we demonstrate how quantum algorithms and hardware can be employed to model stochastic chemical kinetics as described by the CME using the Schl\"ogl Model of a trimolecular reaction network as an illustrative example. To ground our study of the performance of our quantum algorithms, we first determine a range of suitable parameters for constructing the stochastic Schl\"ogl operator in the mono- and bistable regimes of the model using a classical computer and then discuss the appropriateness of our parameter choices for modeling approximate kinetics on a quantum computer. We then apply the Variational Quantum Deflation (VQD) algorithm to evaluate the smallest-magnitude eigenvalues, $\lambda_0$ and $\lambda_1$, which describe the transition rates of both the mono- and bi-stable systems, and the Quantum Phase Estimation (QPE) algorithm combined with the Variational Quantum Singular Value Decomposition (VQSVD) algorithm to estimate the zeromode (ground state) of the bistable case. Our quantum computed results from both noisy and noiseless quantum simulations agree within a few percent with the classically computed eigenvalues and zeromode. Altogether, our work outlines a practical path toward the quantum solution of exponentially complex stochastic chemical kinetics problems and other related stochastic differential equations.  
\end{abstract}

\pacs{}
\maketitle

\section{Introduction}
\label{sec:intro}

One of the most relevant yet underexplored applications of quantum computing in chemistry lies in chemical kinetics, which seeks to determine the transition rates and non-equilibrium steady states of a system by solving its the underlying stochastic differential equations (SDEs).\cite{liu_solving_2022} Since analytical solutions are limited to only a few special cases, numerical methods are typically employed to solve SDEs classically. However, due to the rapid expansion of their state space as their number of dimensions and degrees of freedom increase, solving SDEs is extremely challenging. For more exact approaches for solving SDEs such as the Chemical Master Equation (CME), the space complexity scales exponentially with the number of reacting species in the system. Quantum algorithms that can potentially overcome this classical scaling would thus enable the study of the dynamics of the larger, more complex reaction networks often found in biology,\cite{baiardi_quantum_2023, ullah_stochastic_2010} chemical biophysics,\cite{ge_stochastic_2012} and atmospheric science.\cite{tennie_quantum_2023} 

Several quantum algorithms for solving Ordinary Differential Equations (ODEs) and Partial Differential Equations (PDEs) have been developed over the years, including the Quantum Phase Estimation (QPE)\cite{dobsicek_arbitrary_2007} and the Harrow-Hassidim-Lloyd (HHL) algorithms.\cite{harrow_quantum_2009} Additionally, numerous quantum-classical hybrid algorithms suitable for Noisy Intermediate-Scale Quantum (NISQ)\cite{cao_quantum_2019, cerezo_variational_2021, fedorov_vqe_2022} computers have emerged in recent years. These algorithms typically utilize shallow quantum circuits, which are less prone to noise, and allocate part of the computation to a classical computer. Examples of such algorithms are Variational Quantum Eigensolvers (VQEs),\cite{peruzzo_variational_2014, 
mcclean_theory_2016, romero_strategies_2018} the Variational Quantum Deflation (VQD) algorithm,\cite{higgott_variational_2019} and the Variational Quantum Singular Value Decomposition (VQSVD) algorithm.\cite{wang_variational_2021} All of these algorithms involve classical optimization of a set of free parameters that define the variational ans\"atz (or, in other words, that are used to construct the parameterized quantum circuits needed to represent the quantum state). These parameters are then iteratively fed back into the quantum circuit until convergence is achieved. Variational quantum algorithms have been utilized to estimate the ground and excited states of small molecules such as $\ce{H_2}$,\cite{kandala_hardware-efficient_2017, tilly_computation_2020, wen_full_2024} $\ce{LiH}$,\cite{kandala_hardware-efficient_2017, tilly_computation_2020, wen_full_2024} $\ce{HF}$,\cite{kandala_hardware-efficient_2017} and $\ce{BeH_2}$,\cite{powers_using_2023} as well as quantum magnets.\cite{kandala_hardware-efficient_2017} Recently, VQE was employed as a density matrix embedding solver in an \textit{ab initio} simulation of strongly correlated materials.\cite{cao_ab_2023} 

Although such variational quantum algorithms have existed for over a decade, the majority of research efforts have been focused on solving the Schr\"odinger Equation (SE). \cite{clinton_hamiltonian_2021, benenti_quantum_2008, huggins_unbiasing_2022, smart_verifiably_2024} However, many types of non-Schr\"odinger differential equations are related and can be transformed into one another under different approximations, opening up the possibility of leveraging quantum computers to also accelerate the solution of these equations, as has been discussed in recent seminal works.\cite{tennie_solving_2024, gnanasekaran_efficient_2023, babbush_exponential_2023, kubo_variational_2021} In the context of stochastic processes, nearly all other types of stochastic differential equations can be transformed into the form of a Fokker-Planck Equation (FPE),\cite{chow_path_2015, gardiner_handbook_2009} a partial differential equation that describes the evolution of a system's state variables in the presence of stochastic fluctuations. This includes the Master Equation, the Chapman-Kolmogorov Equation, and both Stratonovich's and It\^o's Stocahastic Differential Equations.\cite{risken_fokker-planck_1996, gardiner_handbook_2009}

In theory, the Chemical Master Equation is the ideal choice for modeling continuous-time stochastic processes in chemistry. The CME describes the evolution of the probability distribution of a system with discrete states in continuous time. The eigenvalues of the CME matrix provide information about transition rates and state lifetimes, while the corresponding eigenfunctions provide information about the state of the system, e.g., the long-term behavior and stability of a system can be described by the eigenfunction corresponding to the lowest eigenvalue.\cite{vellela_stochastic_2009, macnamara_stochastic_2008} The CME's solutions are typically considered exact since it considers all possible states.  The CME is therefore a very resource-intensive approach due to the exponential explosion of the size of its state space with respect to the number of reacting species: a chemical system with $N$ molecules, $R$ reactions, and $S$ reacting species requires an $[R(N+1)]^S \times [R(N+1)]^S$ stochastic matrix to fully capture all possible states. The CME can be approximated by an FPE-type equation in the rapid reaction rate limit in which chemical reaction rates are much faster than diffusion rates.\cite{sjoberg_fokkerplanck_2009, gillespie_chemical_2000} However, solving the CME directly is computationally prohibitive for the vast majority of applications outside the small population limit, even when numerical approximations such as uniformization and the Krylov subspace methods are applied.\cite{wolf_solving_2010, dinh_understanding_2016} Finding more efficient ways of solving the CME is still an active area of research in mathematics and computational science.\cite{ocal_model_2023}

Surprisingly, the CME has garnered little to no attention from the quantum computing community despite its significance in chemistry, biology, physics, and engineering and the computational hardships associated with solving it classically.\citep{gupta_comparison_2014, sukys_approximating_2022} The closest effort to address the CME and chemical stochastic differential equations in general using quantum computing is perhaps the work by Pravatto $\textit{et al.}$, in which the authors estimate the lowest non-zero eigenvalue of a Fokker–Planck-Smoluchowski operator describing the isomerization process in a chain of molecular rotors using a VQE approach.\cite{pravatto_quantum_2021} This work's results confirm an increase in errors in the presence of noise and for larger basis sets, as also observed in quantum chemistry simulations. \cite{oliv_evaluating_2022, saib_effect_2021}  However, there are fewer mitigation strategies available for classical chemical systems, as most mitigation strategies are typically designed with quantum chemistry in mind. For instance, it is well known that the technique used to map and encode a molecular Hamiltonian into qubits can significantly affect the performance of a quantum algorithm, yet most of the developed encoding schemes, including the Jordan–Wigner Transformation\cite{jordan_uber_1928, ovrum_quantum_2007,tranter_comparison_2018}, Bravyi–Kitaev Transformation,\cite{bravyi_fermionic_2002, seeley_bravyi-kitaev_2012} and Trotterization,\cite{trotter_product_1959,wu_polynomial-time_2002,shende_recognizing_2004, kluber_trotterization_2023} are designed for encoding and simulating quantum mechanical systems.\cite{li_variational_2019} Similarly, more advanced versions of VQE that involve the adaptive, gradual growth of an ans\"atz, e.g., ADAPT-VQE,\cite{grimsley_adaptive_2019} are tailored to molecular Hamiltonians starting from a Hartree-Fock approximation. This lack of capabilities can be partially attributed to a limited appreciation of the potential of quantum algorithms for solving problems in classical chemistry, further compounded by the absence of demonstrations in the literature.  

Our work aims to address some of these shortcomings by extending the application of quantum computing to classical chemical kinetics and analyzing the benefits it may bring and challenges it may face. We present a detailed analysis of the classical and quantum solutions to the Schl\"ogl model,\cite{schlogl_chemical_1972} a paradigmatic chemistry model describing the dynamics of a trimolecular autocatalytic process. The Schl\"ogl model describes an inherently stochastic chemical reaction network and is therefore most accurately described by a CME. Owing to device constraints such as limited gate fidelity, gate crosstalk, and the ability to support only a few physical qubits,\citep{cerezo_variational_2021} solutions to the Schl\"ogl model are possible only for a limited set of system parameters on current-NISQ computers. We discuss the suitable range of parameters for simulating this model and the rationale behind our parameter choices. We then apply VQD to determine the first two eigenvalues of our system in order to estimate rates and waiting times, and use a combination of QPE and VQSVD to approximate its steady state solution. We run our numerical simulations in Qiskit using a local simulator, the QASM simulator backend and {\tt ibm\_brisbane} with and without noise, and show that, for the first eigenvalue, the basis employed can be halved without sacrificing accuracy. Our results for both the eigenvalues and the non-equilibrium steady-state are in good agreement with classical results: notably, we obtain exact quantum-computed eigenvalues for the 2- and 3-qubit operator of the Schl\"ogl model and root mean-squared (RMS) deviation errors as low as 3\% for the quantum-computed non-equilibrium steady-state. Our work showcases the potential for quantum computing to solve classical chemistry problems and highlights the need for more efficient mapping and transformation techniques that are tailored to classical operators. To our knowledge, this is the first work to demonstrate the quantum simulation of classical chemical kinetics on near-term quantum hardware. 

The rest of the paper is organized as follows: In Sec. \ref{Chemical kinetics}, we provide relevant background regarding the Chemical Master Equation and Schl\"ogl model before describing the techniques we used to solve them on both classical and quantum hardware in Secs. \ref{sec:methods} and \ref{Implementation}. In Sec. \ref{sec:results}, we benchmark our results on both the noiseless (Qiskit) statevector simulator and {\tt ibm\_brisbane} against classical results, highlighting what simplifications can accurately be made along the way. Lastly, we contextualize our numerical results in Sec. \ref{Discussion} and conclude with a future outlook in Sec.  \ref{Conclusions and outlook}.

\begin{figure*}
    \includegraphics[width=0.9\textwidth]{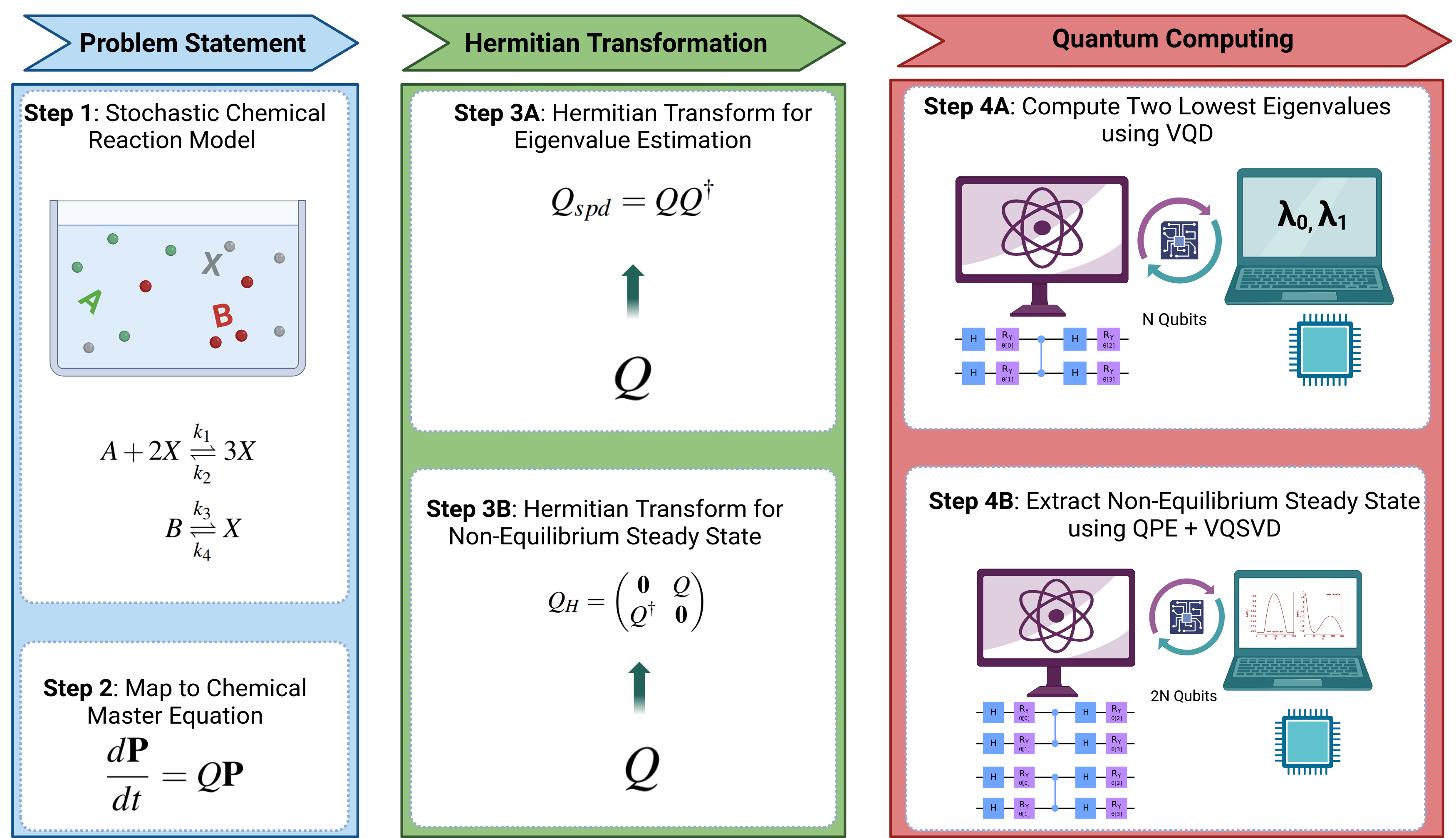}
    \caption{Overview of the workflow presented in this manuscript. In \textbf{Step 1}, a stochastic chemical reaction network is given as the problem statement. The reaction network is then mapped to a Chemical Master Equation with stochastic matrix $Q$ in \textbf{Step 2}. In \textbf{Step 3}, the stochastic matrix is converted to Hermitian form. Finally, in \textbf{Step 4}, the two lowest eigenvalues and the non-equilibrium steady state are computed using variational quantum algorithms.}
    \label{fig:graphical_abstract}
\end{figure*}

\section{Stochastic Chemical Kinetics}
\label{Chemical kinetics}
Deterministic chemical kinetics are typically described using rate equations that predict the changes in concentrations of reacting species with time. However, in the presence of noise, e.g., due to fluctuations in the counts of particles for systems involving small molecular populations such as biochemical processes like DNA transcription, regulation,\cite{blake_noise_2003, kulasiri_chemical_2021, fedoroff_small_2002} and apoptosis,\cite{eissing_bistability_2004} a stochastic description of the reaction kinetics is required. The CME provides the most accurate but also the most computationally demanding approach to solving stochastic dynamics. This is due to the CME's high-dimensional state space, i.e., each chemical species considered in simulations adds a dimension to the CME.\cite{einkemmer_low-rank_2024, wolf_solving_2010} Approximations can be made using the Chemical Fokker-Planck (CFPE) and the Chemical Langevin Equations (CLE),\citep{gillespie_chemical_2002} and stochastic numerical approaches have been developed such as Gillespie's stochastic simulation algorithm (SSA).\cite{gillespie_exact_1977} However, like any Monte Carlo-based simulation method, the sampling error in SSA can be very challenging to estimate and its convergence may be very slow for large systems or rate constants.\cite{macnamara_stochastic_2008} Meanwhile, the CFPE and CLE can yield inaccurate results outside of the thermodynamic limit and for systems with small volumes, e.g., biochemical reactions inside a cell.\cite{gaveau_master_1997, grima_how_2011, van_kampen_validity_1981, gillespie_chemical_2000} Our work presents a proof-of-concept alternative to solving the CME on near-term quantum hardware.

\subsection{Chemical Master Equation}
The CME is an ODE describing the evolution of a continuous time, discrete space Markov process. In the context of biochemical reactions, it describes a system of $N$ spatially homogeneous molecular species $\left\{ S_1, S_2,..., S_N\right\}$ interacting through $M$ chemical reaction channels $\left\{R_1, R_2, ..., R_M \right\}$ at a constant volume and in thermal equilibrium. If we denote the number of $S_i$ molecules in the system at time $t$ by $X_i(t)$, where $i = (1, 2, ..., N)$, the molecular population vector is given by $\mathbf{X}(t) \equiv (X_1(t), X_2(t), ..., X_N(t))$, where $\mathbf{X}(t)$ changes stochastically due to the presence of noise in the system. Rigorous derivations of the CME\cite{gillespie_rigorous_1992, gillespie_chemical_2000, gillespie_chemical_2002} show that, if the reacting species are confined to a specific volume, kept well-stirred, and held at a constant equilibrium temperature, the probability for one reaction $R_j$ to occur in the system in the next infinitesimal time interval $[t, t + dt)$ is given by
\begin{equation}
\label{eq:probability}
P(\mathbf{x},t | \mathbf{x}_0,t_0) = a_j(\mathbf{x}) dt,
\end{equation}
where $a_j$ is the propensity function, $j = (1, 2, ..., M)$, and $\mathbf{X}(t) = \mathbf{x}$ is an $N$-dimensional {\it Markov jump process}, i.e., $\mathbf{X}(t)$ executes a {\it random walk} in the $N$-dimensional state space. Thus, for each $j$, a state-change vector 
\begin{equation}
\label{eq:state-change}
    \mathbf{v}_j \equiv (v_{j1}, v_{j2},...,v_{jN})
\end{equation}
is defined using the change in the number of $S_i$ molecules as a result of one reaction $R_j$. Given an initial probability $P(\mathbf{X}_0, t_0)$ and that $t \geq t_0$, the time evolution of the probability function satisfies Equations \ref{eq:probability} and \ref{eq:state-change} and is given by the CME\cite{gillespie_chemical_2000}
\begin{equation}
\begin{split}
\frac{d}{dt}P(\mathbf{x},t | \mathbf{x}_0,t_0) = \sum_{j=1}^{M} & \left[ a_j(\mathbf{x} - \mathbf{v}_j) P(\mathbf{x} - \mathbf{v}_j, t | \mathbf{x}_0,t_0) \right. \\
& \left. - a_j(\mathbf{x}) P(\mathbf{x}, t |\mathbf{x}_0,t_0) \right] .
\end{split}
\end{equation}
The CME has been rigorously shown to be an exact description of the microscopic physics of a system as it considers every possible state by taking into account all the possible reactions. For sufficiently large molecular populations, the CME can be approximated by a CLE, CFPE, or a reaction rate equation (RRE), which all assume a continuous Markov process. For a detailed discussion of the relationship between the CLE, CFPE, and RRE to the CME, we direct the reader to Refs. \citenum{gillespie_chemical_2000, gillespie_chemical_2002, gillespie_rigorous_1992}.

\subsection{The Schl\"ogl Model}
\label{sec:schlogl}
The Schl\"ogl model,\citep{schlogl_chemical_1972} named after its author, F. Schl\"ogl, is a chemical reaction network that, in its stochastic form, exhibits bistability. The model was first developed to understand non-equilibrium phase transitions in well-stirred chemical reactions by analyzing the steady and unstable states of a system with respect to the concentrations of the reacting species.

Due to the inherent randomness of the collisions of the molecules in a stirred mixture, the kinetics of a well-stirred chemical composition can  be modeled using stochastic differential equations. If the system comprises a considerable number of particles, or in statistical mechanics terminology, approaches the thermodynamic limit of infinite molecular populations, the Langevin and Fokker-Planck stochastic differential equations can be used. However, the CME is more suitable if the system is characterized by small molecular populations, as is the case in cellular biological systems and biochemical processes.\citep{gillespie_chemical_2002,ilie_numerical_2009} In the Schl\"ogl model, the concentrations of some of the species are kept constant while the concentrations of other species are allowed to change. The species with variable concentrations are referred to as dynamic species. The Schl\"ogl model can be deterministic or stochastic depending on whether the concentration of the dynamic species is allowed to fluctuate randomly or according to a predictable pattern. If we let $A$ and $B$ signify the constant species and $X$ the dynamic species, a simple Schl\"ogl model will be given by the reaction network\citep{schlogl_chemical_1972, vellela_stochastic_2009}
\begin{subequations}
\begin{align}
 A + 2 X &\underset{k_2}{\stackrel{k_1}{\rightleftharpoons}} 3X\\
 B &\underset{k_4}{\stackrel{k_3}{\rightleftharpoons}} X
\end{align}
\end{subequations}
where $k_1$, $k_2$, $k_3$, and $k_4$ are rate constants. 
The deterministic Schl\"ogl model can be obtained using the law of mass action,  \citep{qian_kinetic_2021} which posits a direct relationship between the rates of a chemical reaction and the concentrations of its reacting species. Let $a$ and $b$ be the concentrations of the static species and $x$ be the concentration of the dynamic species. The deterministic Schl\"ogl model for the reactions above can be written in the form of an ordinary differential equation\citep{vellela_stochastic_2009}
\begin{equation}
    \label{eq:shcl1}
    \frac{dx}{dt} = k_1 a x^2 -k_2 x^3 - k_4 x + k_3 b.
\end{equation}
The stochastic Schl\"ogl model is described by a CME with infinitely-coupled ODEs truncated at some finite number $N$ for numerical purposes. It can be derived by first defining the number of molecules of $A$, $B$, and $X$ in a fixed volume $V$. Let $n_A$ and $n_B$ be the number of $A$ and $B$ molecules, and $n_X(t)$ be the number of $X$ molecules at time $t$. From here, one can define a probability distribution function $P_n(t)$ that gives the probability of finding $n$ number of $X$ molecules at time $t$. The stochastic Schl\"ogl model will then be governed by a {\it birth-death} Markov process with the following CMEs:
\begin{subequations}
\label{eq:master}
\begin{align}
\frac{d P_0(t)}{dt} &= \mu_1P_1 - \kappa_0P_0,\\
\frac{d P_n(t)}{dt} &= \kappa_{n-1} + \mu_{n+1} P_{n+1} - (\kappa_n + \mu_n)P_n,
\end{align}
\end{subequations}
where $n = [1,\infty)$, i.e., the stochastic model constitutes an infinite system of coupled ordinary differential equations. Here, $\kappa_n$ and $\mu_n$ are known as the birth and death rates of the Markov process, respectively. They are the recurrence relations used to construct the stochastic matrix and  may be written in terms of the deterministic rate constants $k_1$, $k_2$, $k_3$, and $k_4$, and the volume $V$ of the system:  \citep{vellela_stochastic_2009}
\begin{subequations}
\label{eq:bdr}
\begin{align}
\kappa_n &= \frac{ak_1n(n-1)}{V} + bk_3V,\\
 \mu_n &= nk_4 + \frac{k_2n(n-1)(n-2)}{V^2}.
\end{align}
\end{subequations}

\subsection{Discretization Procedure}
\label{Matrix construction}
\begin{figure*}
     
     \subfloat[Monostable steady states.]{{\includegraphics[width=0.45\textwidth]{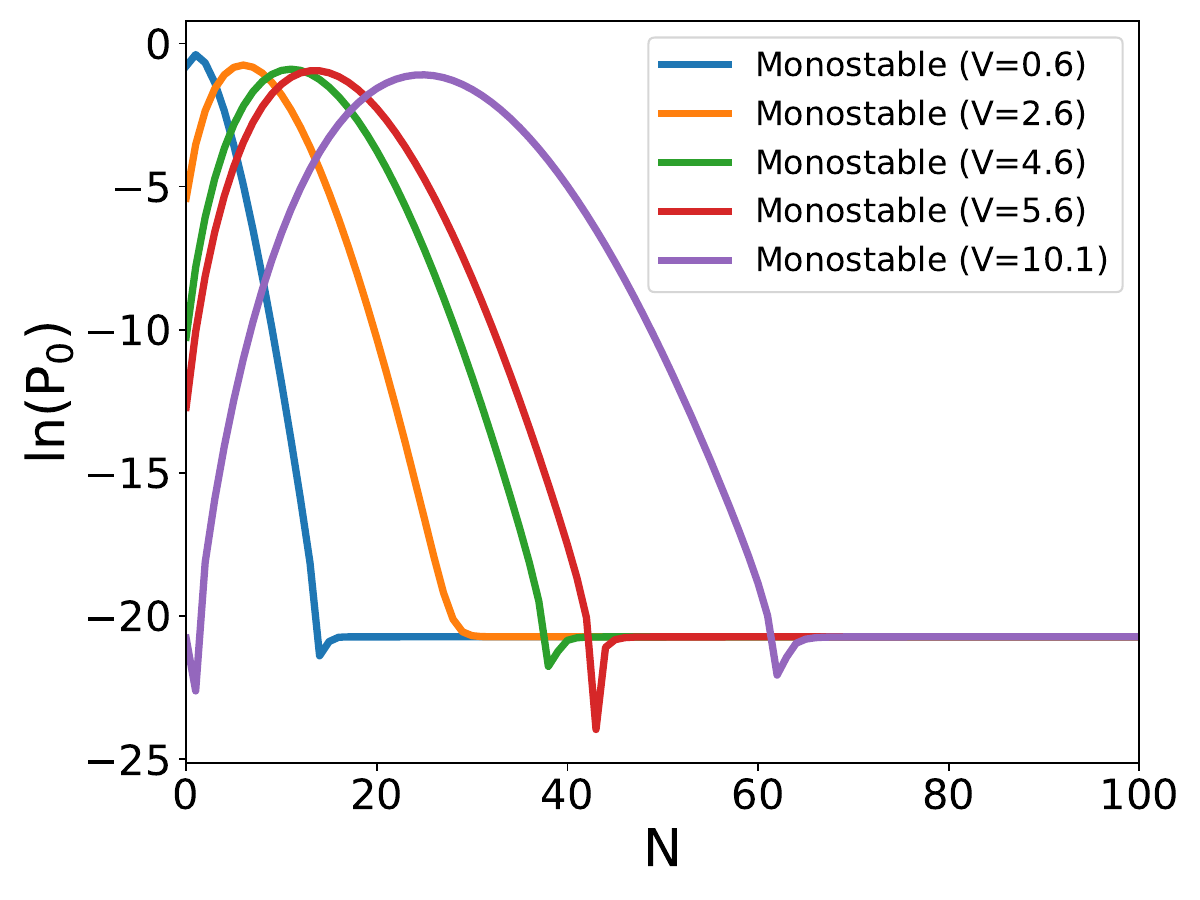} }}%
   %  \qquad
     \subfloat[Bistable steady states.]{{ \includegraphics[width=0.45\textwidth]{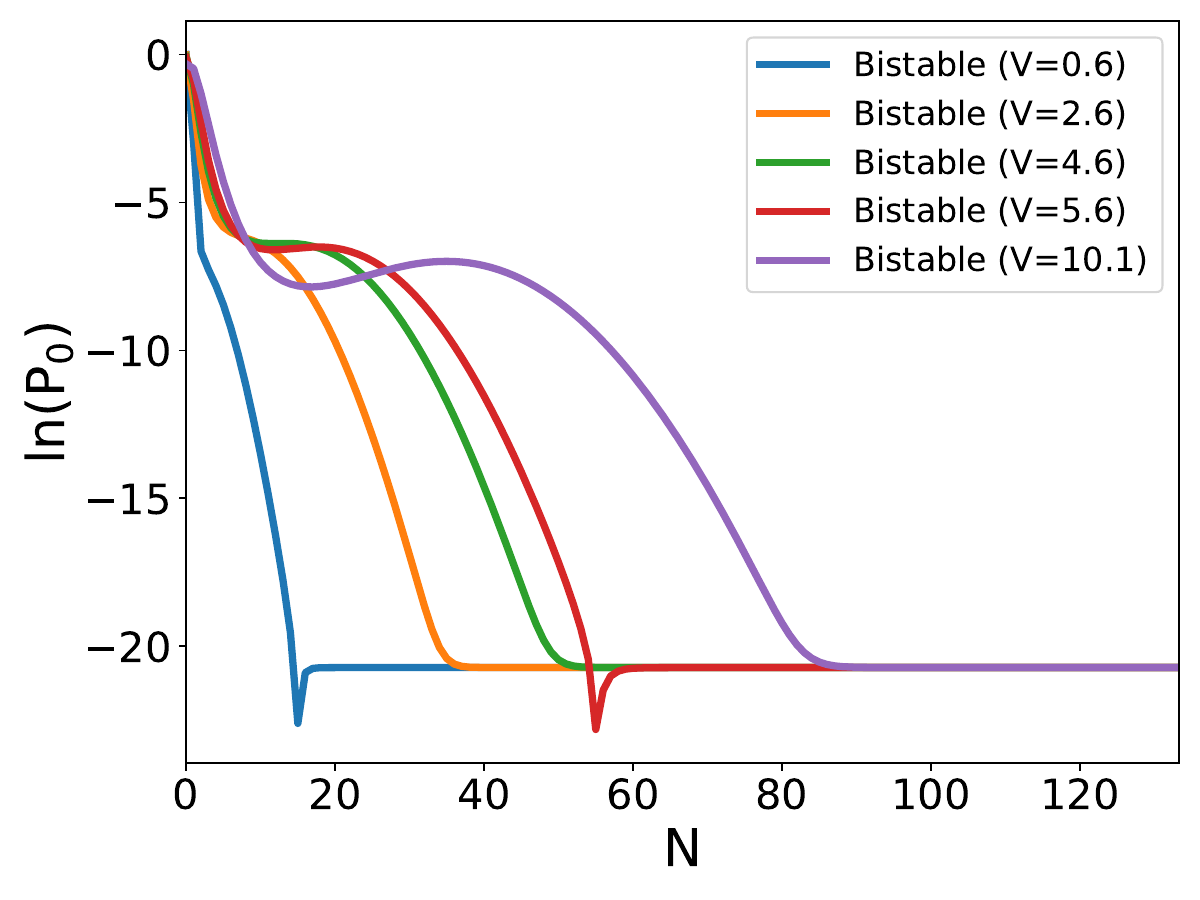} }}%
     \caption{Monostable and bistable steady states of the Schl\"ogl model as a function of the basis size, $N$, for different volumes, $V$. The monostable steady state is characterized by a Poisson distribution with a single peak while the bistable case is characterized by a deformed Poisson-like distribution with two peaks separated by a trough. All plots are given on a log scale.}
     \label{fig:multi-steadystates}%
 \end{figure*}

Discretizing the Schl\"ogl model is relatively straightforward since the CME is a discrete differential equation. All we need to do is cast Equation \ref{eq:master} into a matrix by taking into account the recurrence relations given by Equation \ref{eq:bdr}. The goal is to obtain the stochastic matrix, $Q$, that describes the Schl\"ogl model through the following stochastic differential equation:
\begin{equation}
\label{eq:stochastic_diff}
    \frac{d \mathbf{P}}{dt} = Q \mathbf{P}
\end{equation}
where each component of the vector $\mathbf{P}$ represents the probability for each state. The solution $\mathbf{P}(t)$ can then be given in terms of the eigenvectors and eigenvalues of $Q$ as follows:
\begin{equation}
    \mathbf{P}(t) = c_0 \mathbf{v}_0 e^{\lambda_0 t} + c_1 e^{\lambda_1 t} \mathbf{v}_1 + ...,
\end{equation}
where $(\lambda_0, \lambda_1, ... )$ are the eigenvalues and $(\mathbf{v}_0, \mathbf{v}_1, ...)$ are the {\it right} eigenvectors of $Q$. The vector $\mathbf{v}_0$ corresponds to the lowest eigenvalue and represents the steady state solution (i.e., $\frac{d\mathbf{P}}{dt} = 0$) of the system. The stochastic matrix $Q$ has an infinite form\cite{vellela_stochastic_2009}
\begin{equation}
    \label{eq:schl3}
     Q =\begin{pmatrix}
-\kappa_0 & \mu_1 & 0 & \cdots \\
\kappa_0 & -\kappa_1-\mu_1 & \mu_2 & \cdots \\
0 & \kappa_1 & -\kappa_2-\mu_2 & \cdots \\
\vdots &\vdots & \vdots &\cdots 
\end{pmatrix},
\end{equation}
but, in practice, must be truncated at a finite number, $N$, such that $\kappa_N = 0$ and $\mu_{N+1} = 0$. $Q$ has real and negative eigenvalues starting from $\lambda_0 = 0$. Since $\mathbf{v}_0$ corresponds to the zero eigenvalue, it is also known as the zeromode. Figure \ref{fig:multi-steadystates} shows the zeromode of the Schl\"ogl model for various sets of parameters. These steady states give the long-term behavior of the system. The stochastic nature of the Schl\"ogl model can be seen through the existence of a bistable regime when the equilibrium condition given by Equation \ref{eq:equilibrium} is not satisfied (see Figure \ref{fig:multi-steadystates}(b)). This non-equilibrium steady state is characterized by two peaks separated by a trough (an unstable steady state) corresponding to two different states in which the system is likely to be, e.g., the two states in which a cell performs its functions.\cite{vellela_stochastic_2009} As shown in Figure \ref{fig:multi-steadystates}(b), one of the peaks will be more dominant than the other, i.e., will have a higher probability. However, since the trough corresponds to a non-zero value on the $y$-axis, there is a non-zero probability of transitioning to the other steady state, i.e., the non-dominant peak. Given that these {\it functional} states have unequal probabilities, the bistable Schl\"ogl model predicts that the system will spend more time in one state than the other.
%An important quantity one can extract from the Schl\"ogl model is $\lambda_1$---the smallest non-zero eigenvalue of $Q$, which signifies the time spent in one of the functional states. Since $\lambda_1$ represents the transition rate, it is directly related to the waiting time, i.e., time spent in each state. 
An important quantity one can extract from the stochastic matrix $Q$ is its smallest non-zero eigenvalue denoted by $\lambda_1$, which is directly proportional to the transition rate but inversely related to the waiting time (i.e., time spent in a functional state). Let $r^+$ and $r^-$ represent the transition rates from the dominant peak to the non-dominant peak and vice versa, respectively. If we denote the waiting times in the dominant and non-dominant states by $T^+$ and $T^-$, respectively, $\lambda_1$ is defined as\cite{hanggi_bistable_1984}
\begin{equation}
    \lambda_1 = r^+ + r^- = -\frac{1}{2T^+} - \frac{1}{2T^-}.
\end{equation}
In the bistable case, the magnitude of $\lambda_1$ decays exponentially as $V$ increases, which means that for larger volumes, distinguishing between $\lambda_0$ and $\lambda_1$ is more challenging. From Figure \ref{fig:multi-steadystates}(b), it can also be seen that the bistability of the system becomes more apparent as the volume increases. As illustrated in Figure \ref{fig:multi-steadystates}, a larger basis is required to represent steady states with larger volumes. From Figure \ref{fig:multi-steadystates}(b), it can be seen that the bistability becomes more pronounced for larger volumes. The eigenvalues follow a similar trend: as $V$ increases, $\lambda_1$ approaches $\lambda_0$ such that the states become indistinguishable. This results in a steady steady with two peaks.
% %%%%%% figure # 1

\subsection{Computational Complexity of Solving the Chemical Master Equation}
\label{sec:complexity}
The greatest challenge to numerically solving the CME is the rapid expansion of its state space with the number of reacting species $S$, number of reaction channels $R$, and the maximum count of molecules per species $N$. In terms of the above, the total number of possible states in the system is given by
\begin{equation}
    \mathcal{N} = [R(N+1)^S],
\end{equation}
which includes the zero state in which a particular molecule is absent or completely used up. The stochastic matrix representing such a system will therefore have dimensions
\begin{equation}
 \mathcal{M} =   [R(N+1)^S] \times [R(N+1)^S],
\end{equation}
i.e., its dimensionality grows exponentially with the number of reacting species, $S$. Exact diagonalization typically scales as $\mathcal{O}(\mathcal{N}^3)$, where $\mathcal{N}$ denotes the dimension of the matrix in question. Therefore, solving a CME by exact diagonalization may carry a space complexity of $\mathcal{O}((N+1)^{3S})$, which is intractable for most systems encountered in chemistry. The vast majority of numerical approaches for solving the CME therefore strive to alleviate this {\it curse of dimensionality}\cite{bellman_dynamic_1957} by using state space reduction methods such as the Krylov subspace and uniformization methods.\cite{wolf_solving_2010, kulasiri_chemical_2021} Krylov subspace methods involve iteratively growing the state space by finding {\it subspaces} that represent the original system while uniformization approaches attempt to solve a transformed CME with uniform transition rates. Obtaining exact solutions to chemical dynamics problems like the ones highlighted in this work is thus exponentially costly on classical computers, making these systems potential targets of opportunity for quantum algorithms; the techniques described below attempt to take a step in this direction.

\section{Methods}
\label{sec:methods}
\subsection{Hermitian Transformations}
\label{sec:hermitian}
To solve this problem using variational quantum eigensolvers fit for near-term quantum architectures,\citep{cerezo_variational_2021} the matrix \ref{eq:schl3} must be transformed into Hermitian form so that the first and second eigenstates can be distinguished by virtue of them being orthogonal to each other. The Hermitian form of Equation \ref{eq:schl3} can be obtained through the transformation

\begin{equation}
    \label{eq:Qh}
    Q_H = \begin{pmatrix}
\mathbf{0} &Q  \\
Q^{\dagger} &\mathbf{0} 
\end{pmatrix},
\end{equation}
where $\mathbf{0}$ denotes the zero matrix and $Q^{\dagger}$ is the conjugate transpose of $Q$. The matrices \ref{eq:schl3} and \ref{eq:Qh} are connected by a singular value decomposition such that the eigenvalues of Equation \ref{eq:Qh} are the singular values of Equation \ref{eq:schl3}. The singular values are not generally similar to the eigenvalues; however, for matrices that are nearly Hermitian, the singular values will be proportional to the eigenvalues. Equation \ref{eq:Qh} can thus be considered a Hermitian approximation to the non-Hermitian stochastic problem \ref{eq:stochastic_diff}. Figure \ref{fig:classical-eigenvalues} shows the classically computed eigenvalue of the Schl\"ogl model for different values of $V$. We show that the non-Hermitian eigenvalues of the bistable system are a good approximation to the eigenvalues of the original operator. For volumes between 0.1 and 20.1 in steps of 1.0, we obtained a root mean squared error (similar to Equation \ref{RMS deviation}) of 0.15 for $\lambda_1$ between the Hermitian and non-Hermitian operators. In the same volume range, we obtained a coefficient of determination, $R^2$, of 0.95. $R^2$ was computed using
\begin{equation}
    R^2 = 1 - \frac{\sum_i\left(y_i^{\mathrm{classical}}-y_i^{\mathrm{quantum}}\right)^2}{\sum_i\left(y_i^{\mathrm{classical}}-y_{\mathrm{mean}}^{\mathrm{classical}}\right)^2},
\end{equation}
where $y_i^{\mathrm{classical}}$, $y_i^{\mathrm{quantum}}$, and $y_{\mathrm{mean}}^{\mathrm{classical}}$ denote the eigenvalues obtained classically, via quantum algorithms (VQD), and the mean of the classically obtained eigenvalues, respectively. In our quantum simulations, we therefore focus on the eigenvalues of the bistable system and obtain the zeromode (non-equilibrium steady-state) in this regime of the Schl\"ogl model.

\begin{figure}[!t]
     \centering
     \subfloat[\centering Hermitian Schl\"ogl matrix at $V=33$.]{{\includegraphics[width=0.50\textwidth]{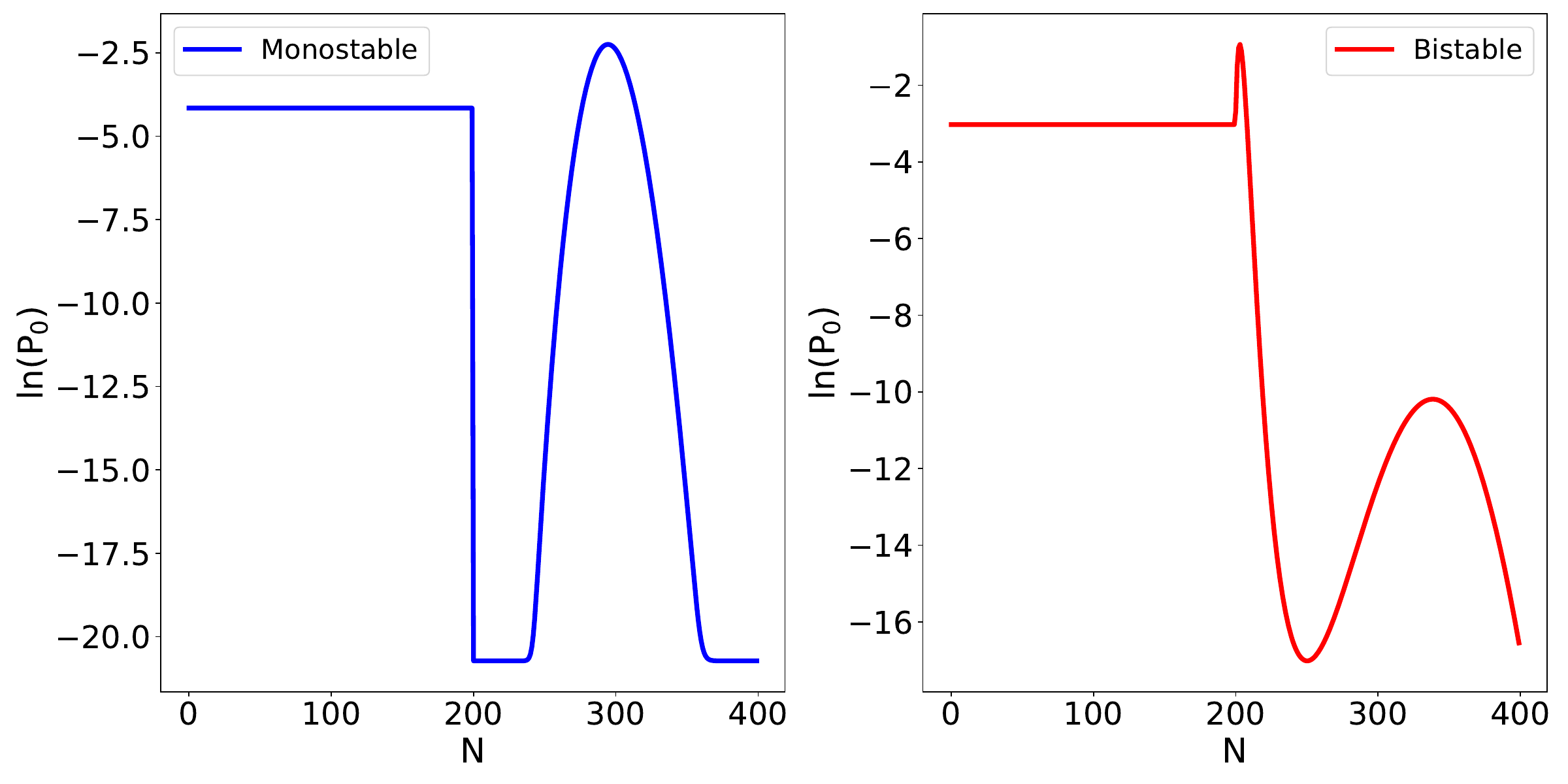} }}%
     
     \subfloat[\centering Non-Hermitian Schl\"ogl matrix at $V=33$.]{{\includegraphics[width=0.50\textwidth]{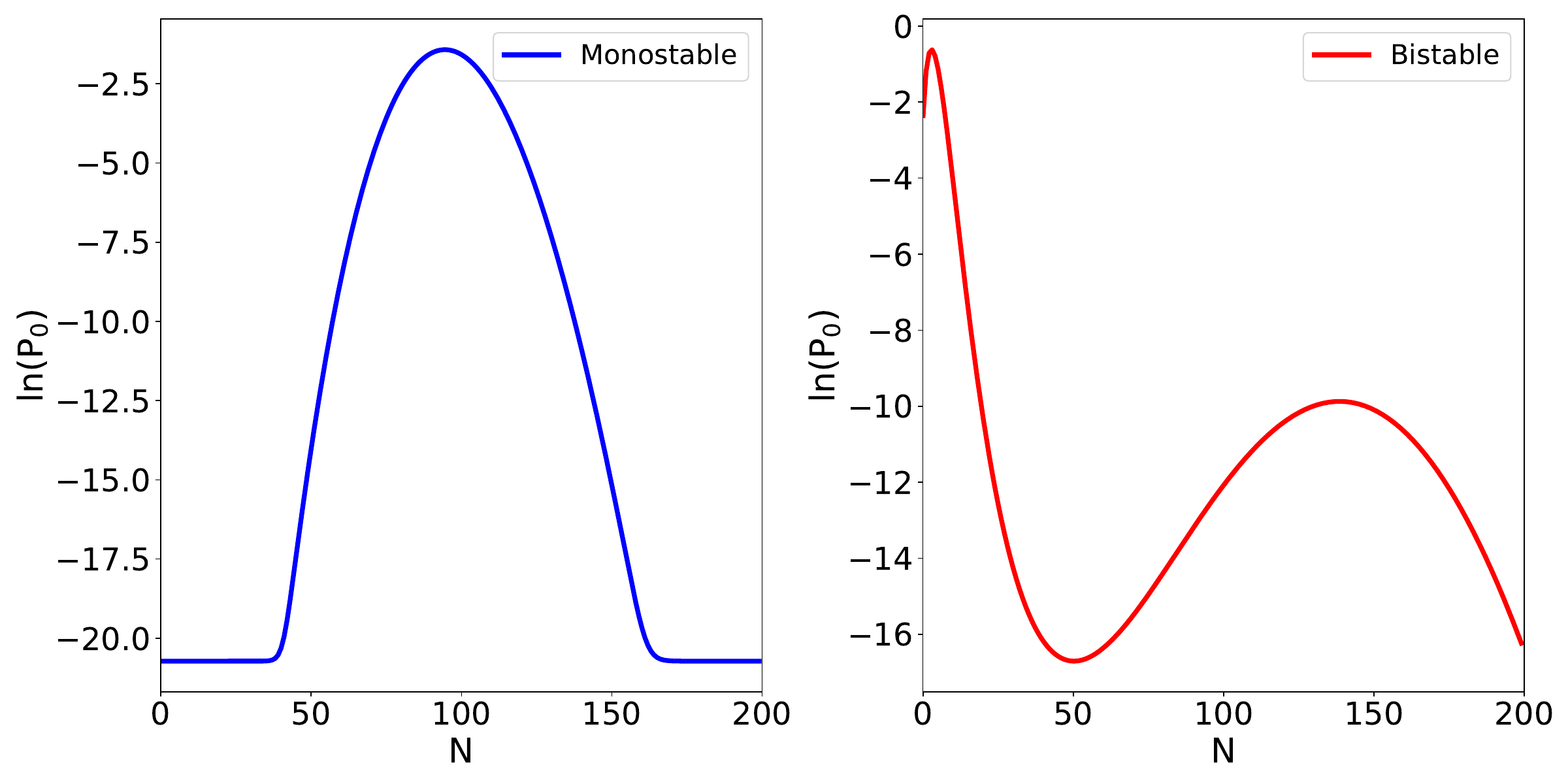} }}%
    \caption{(a) Monostability (left) and bistability (right) of the Schl\"ogl model at $V=33$. The rate constants for both the monostable and bistable systems are as follow: $k_1=3$; $k_2=0.6$; $k_3=0.25$; and $k_4=2.95$. The pump parameters for the monostable state are $a = 0.5$ and $b=29.5$. For the bistable state, $a = b = 1$. \citep{vellela_stochastic_2009} (b) Same as (a), but for a non-Hermitian Schl\"ogl operator.}%
     \label{fig:steady}%
 \end{figure}

Note that Equation \ref{eq:Qh} requires a basis size twice as large as the original operator. Although the eigenvectors of the original operator are essentially conserved in this transformation, the first (left) half of the components in the computed eigenvectors must be discarded as shown in Figure \ref{fig:steady}. The first half of the zeromode (steady state) of the Hermitian operator is an artifact of the $\mathbf{0}$ matrices used in the transformation---only the latter (right) half of the eigenvector components carry information about the steady state. In our quantum simulations, we used this transformation to recover the zeromode of the Schl\"ogl operator using QPE and VQSVD. For the eigenvalue simulations with VQD, we use the Cholesky decomposition

\begin{equation}
\label{eq:Qspd}
    Q_{spd} = Q Q ^ {\dagger},
\end{equation}
which ensures that the Schl\"ogl operator is semi-positive definite. The eigenvalues of $Q_{spd}$ are the squares of the singular values of $Q$ without the negative sign, since $Q_{spd}$ has only non-negative eigenvalues. Provided the initial matrix is not too far from Hermitian, one can therefore still approximate the eigenvalues of $Q$ using Equation \ref{eq:Qspd}. Note that Equation \ref{eq:Qspd} does not conserve the eigenvectors like Equation \ref{eq:Qh}. However, its smaller basis size makes it more ideal for eigenvalue simulations using VQD. Since the eigenvalues using VQD are computed one after the other starting from the smallest to the largest, the operator $Q_H$ has twice as many eigenvalues. Therefore, it would take twice as many iterations to get to a desired eigenvalue compared to using $Q_{spd}$. Perhaps most importantly for this work, we numerically identify a relationship between the zeromodes of $Q_{spd}$ and $Q$. Specifically, the zeromode of $Q_{spd}$ is exactly equal to the {\it left} eigenvector of $Q$ corresponding to $\lambda_0$. This allows us to implement a modified version of the VQD algorithm, VQD-{\tt exact0}, which skips the first state and directly computes $\lambda_1$. Using an exact initial state for the VQD algorithm also allows us to retrieve an exact first excited state of $Q_{spd}$ corresponding to the eigenvalue $\lambda_1$. The exact initial vector of $Q_{spd}$ used in our simulations is the constant state given by

\begin{equation}
    \label{eq:initial_state}
    \mathbf{w}_0 = \underbrace{ \left[\frac{1}{2^N}, \frac{1}{2^N}, \frac{1}{2^N}, ... \frac{1}{2^N} \right]^T }_{\mathrm{Length} = {2^N}},
\end{equation}
where $N$ is the number of qubits needed to encode $Q_{spd}$. Since $\mathbf{w}_0$ has a length of $2^N$, it is normalized to unity by dividing by this length.

% %%%%%% figure # 2
 
\subsection{VQD}
VQD is a variational quantum algorithm that estimates the excited state energies of a particular Hamiltonian.\citep{higgott_variational_2019} In our numerical simulations, we use VQD to compute the lowest and second lowest eigenvalues of the stochastic Schl\"ogl matrix. 

Given a Hermitian operator (or Hamiltonian), $\hat{H}$, VQD takes in as input a specified number $k$ of the eigenvalues of $\hat{H}$. The objective of VQD is to first find a parameter $\lambda_0$ such that the cost function
\begin{equation}
    \label{VQD cost function}
    E(\lambda_0) = \bra{\psi(\lambda_0)} \hat{H} \ket{\psi(\lambda_0)}, 
\end{equation}
is minimized. Once we determine $\lambda_0$ using the variational quantum eigensolver algorithm,\citep{peruzzo_variational_2014, wang_accelerated_2019} we adopt an iterative procedure to find parameters $\lambda_1$ through $\lambda_k$ such that the cost function, 
\begin{equation}
    \label{VQD cost function 2}
    F(\lambda_{x}) =  E(\lambda_{x}) + \sum^{x - 1}_{i = 0}\beta_{i}\abs{\bra{\psi(\lambda_{x})}\ket{\psi(\lambda_{i})}}^{2}, 
\end{equation}
is minimized $\forall ~x \in [1, k]$ (here, $\beta_i \in \mathbb{R}$). We decompose the Hamiltonian $\hat{H}$ into a linear combination of Pauli strings, i.e., $\hat{H} = \sum_{j} c_{j}\hat{P_{j}}$, to facilitate the implementation of VQD on near-term quantum hardware. 

\subsection{QPE}
\label{QPE}
QPE is a quantum algorithm that estimates the eigenvalues of an unitary matrix.\citep{kitaev_quantum_1995} Here, we use QPE as a consistency check, i.e., to verify if the stationary state of the stochastic Schl\"ogl model can be recovered by a quantum algorithm. Furthermore, we use QPE to estimate the minimum eigenvalue (denoted as $\lambda_{\text{min}}$ for the purpose of discussion) of the unitary representation of the stochastic Schl\"ogl matrix. We use $\lambda_{\text{min}}$ in conjunction with VQSVD to obtain an estimate for the non-equilibrium steady-state of the Schl\"ogl model (see Sec. \ref{VQSVD} for more details).  

\subsection{Non-Equilibrium Steady State from VQSVD}
\label{VQSVD}
We use VQSVD to obtain an estimate for the non-equilibrium steady-state (or zeromode) of the stochastic (bistable) Schl\"ogl matrix. One amongst the several goals of singular value decomposition (SVD) is to examine the null space of any given matrix. SVD seeks to find a decomposition of the form $U D V^{\dagger}$ for some arbitrary matrix $T$, where $U$ and $V$ contain the left and right singular eigenvectors of $T$, respectively, and $D$ is a diagonal matrix that encodes information about the singular values of $T$. 

Current-NISQ devices are capable of supporting only a limited number of physical qubits. One designs variational quantum algorithms by keeping such constraints in mind, i.e., a limited gate fidelity and the ability to support only a few physical qubits on NISQ hardware.\citep{wang_variational_2021} To this end, we employ VQSVD,\citep{wang_variational_2021} a variational quantum algorithm that seeks to implement SVD via an optimization procedure. A detailed discussion of the implementation of VQSVD is beyond the scope of this work (we direct the reader to Ref. \citenum{wang_variational_2021} for more details).

\textit{Finding the non-equilibrium steady-state}---We compute the non-equilibrium steady-state of the stochastic Schl\"ogl model following an eigenvalue equation treatment of the zeromode. We briefly outline the steps involved in obtaining an estimate for the non-equilibrium steady-state, as also illustrated in Fig.~\ref{fig:graphical_abstract}, below: 

\begin{itemize}
    \item Construct a finite volume discretization (i.e., a matrix representation) for the Schl\"ogl operator using the procedure outlined in Sec. \ref{Matrix construction}. Thereafter, apply a block-diagonal transformation to this matrix (we denote this as $Q_H$ for the purpose of discussion; also see Sec. \ref{Matrix construction}). 
    \item Construct the unitary representation of the Hermitian matrix $Q_H$ (i.e., $U = e^{-i Q_H}$). As discussed in Sec. \ref{sec:hermitian}, this transformation preserves the eigenvector spectrum of the Schl\"ogl operator.
    \item Search the null space of the matrix $U - \lambda_{\text{min}} I_{d \times d}$ using VQSVD to obtain an estimate for the non-equilibrium steady-state. As mentioned earlier, this realization follows from the fact that $U$ and $Q_H$ share the same eigenvector spectrum as they are connected via a unitary transformation. We use QPE to estimate $\lambda_{\text{min}}$. 
\end{itemize}

\section{Implementation}
\label{Implementation}

We implement VQD, QPE, and VQSVD using the Qiskit platform.\citep{treinish_qiskitqiskit-metapackage_2023} The {\tt TwoLocal} ans\"atz was used in all of our VQD simulations. For the 2- and 3-qubit operators, we used the 1- and 2- circuit repetitions of $R_Y$ gates with linear CNOT entanglements, respectively. For the 4-qubit operator, similar ans\"atz\"e labeled as A and B, were used with 4- and 5- circuit repetitions, respectively. The number of circuit parameters ranged from 4 (for the 2-qubit operator) to 24 (for ans\"atz B). The classical optimization of the parameters was performed using the Limited-Memory Broyden-Fletcher-Goldfarb-Shanno Bound (L-BFGS-B)\cite{zhu_algorithm_1997} algorithm available in the Qiskit package. All VQD implementations were noiseless using local simulators of Qiskit's {\tt Estimator}, {\tt Sampler}, and {\tt StateVector} classes. For QPE, we ran numerical experiments on {\tt ibm\_brisbane}, a publicly-available 127-qubit quantum processing unit (QPU). We determined the number of qubits used to construct the query qubit register by $n = \log_2(d)$, where $d$ denotes the dimension of $\hat{U}$, the unitary representation of the block diagonal form of the stochastic Schl\"ogl matrix (we denote this as $Q_H$ in the preceding sections). To construct $\hat{U}$, we performed the unitary transformation $\hat{U} = e^{-i Q_H}$ (the eigenvector spectrum of $Q_H$ is preserved under this unitary transformation). We used QPE to estimate the minimum eigenvalue of $\hat{U}$.

We set the number of precision qubits used to construct the precision qubit register equal to seven across all experimental runs. We set the optimization level of the QPE circuit equal to 1 throughout to allow for minimal optimization of the QPE circuit.
%(i.e., we still retain most of the relevant gates in the QPE circuit to facilitate a useful noise analysis). 

Next, we employ VQSVD\citep{noauthor_release_nodate} (implemented on the PaddlePaddle Deep Learning platform\citep{noauthor_paddlepaddle_nodate, bi_paddlepaddle_2022}) to obtain an estimate for the zeromode of the stochastic Schl\"ogl matrix. We use the hardware-efficient $R_Y - R_Z$ ans\"atz, which consists of single-qubit rotation gates and two-qubit entangling CNOT gates (we direct the reader to Ref. \citenum{wang_variational_2021} for more details regarding the ans\"atz architecture). We use the Adam optimizer to perform a classical optimization of the variational parameters. We choose the weights to run VQSVD per the procedure outlined in Ref. \citenum{wang_variational_2021}. Further details pertaining to the setup and implementation of VQSVD may be found in the Supplementary Materials. 

% To interleave DD within the QPE circuit, we followed the procedure outlined in Ref. \citenum{niu_effects_2022}. We note that the DD sequences are applied to the entire QPE circuit, i.e., both the precision and query qubit registers when qubits are idle. We performed an extensive analysis of the performance of the QPE circuit in the presence of noise by applying a number of well-known DD sequences.\citep{hahn_spin_1950, carr_effects_1954, biercuk_experimental_2009, ali_ahmed_robustness_2013, paz-silva_dynamical_2016, alvarez_iterative_2012} To implement the Uhrig-X and Uhrig-Y DD sequences, we inserted a total of eight $X$ and $Y$ single-qubit gates, respectively. 

\section{Numerical results}
\label{sec:results}

\subsection{Parameter Selection}
\label{Parameter selection}
\paragraph{Rate Constants} The Schl\"ogl model can exhibit monostability or bistability depending on the number of species and parameters chosen. Information about the stability of the system can be directly obtained from the steady state solution or the zeromode, i.e., the eigenvector corresponding to the zero eigenvalue (see Figure \ref{fig:multi-steadystates}). The monostable Schl\"ogl model is characterized by a Poisson distribution with a single peak corresponding to the concentration of the dynamical species $x$ required to achieve chemical equilibrium.\cite{heuett_grand_2006, gardiner_handbook_2009} To achieve chemical equilibrium, {\it chemical detailed balance} must be satisfied, i.e., the forward fluxes $J^{+}_i$ must equal the backward fluxes $J^{-}_i$ across the entire reaction network ($J^{+}_i = J^{-}_i$). The equilibrium condition for the Schl\"ogl model will then be given by
\begin{equation}
\label{eq:equilibrium}
    \frac{R_{\rightarrow}}{R_{\leftarrow}} = \frac{k_1 k_4 a}{k_2k_3b} = 1,
\end{equation}
where $R_{\rightarrow}$ and $R_{\leftarrow}$ denote the forward and backward reaction rates in the system, respectively. 

On the other hand, the bistable Schl\"ogl model signifies a departure from equilibrium and is characterized by a deformed Poisson-like distribution with two peaks. Each peak corresponds to a stable steady state, with an unstable steady state (the trough) between them. Since the bistable system is only achieved when the equilibrium condition given by Equation \ref{eq:equilibrium} is not satisfied, each steady state (peak) is actually a non-equilibrium steady state (NESS). Consequently, the bistable system can oscillate and evolve between two NESSs since there is a non-zero probability of moving from one stable state to the other. In principle, any set of parameters that satisfy (violate) Equation \ref{eq:equilibrium} will result in a monostable (bistable) solution. In this work, we chose a set of parameters that illustrated a clear distinction between the monostable and bistable regimes and that had been previously studied analytically. Following Ref. \citenum{vellela_stochastic_2009}, the rate constants used throughout this work are as follows: $k_1=3$; $k_2=0.6$; $k_3=0.25$; and $k_4=2.95$. To satisfy Equation \ref{eq:equilibrium} and achieve monostability, we set $a=0.5$ and $b=29.5$. The bistable case was realized using $a=b=1$. Note that varying $a$ and $b$ while keeping other parameters constant would shift the relative heights of the NESS in Figure \ref{fig:multi-steadystates}(b). Since the birth and death rates given by Equation \ref{eq:bdr} are scaled by $V$, the system volume is also critical in describing bistability.

\paragraph{System Volume} Figure \ref{fig:classical-eigenvalues} shows the effects of changing the volume on the eigenvalues of the monostable and bistable Hermitian and non-Hermitian Schl\"ogl models. For the bistable system, the Hermitian and non-Hermitian results for $\lambda_1$ vary only slightly compared to the monostable case. For larger volumes, the bistable eigenvalues are almost indistinguishable due to the exponential decay of $\lambda_1$---both the Hermitian and non-Hermitian $\lambda_1$ values approach zero when $V$ is very large. For example, setting $V=100$ gives $\lambda_1 \approx 10^{-10}$, while setting $V =1$ yields $\lambda_1 \approx 1.5$. In order to get a first excited state, i.e., $\lambda_1$, that is sufficiently distinguishable from the ground state, $\lambda_0$, the volume must satisfy: $V \rightarrow 0$ (see Figures \ref{fig:original-eigenvalues} and \ref{fig:eig-comparison}). Nevertheless, the Hermitian approximations discussed in Section \ref{sec:hermitian} can provide a reasonable estimate of the Schl\"ogl eigenvalues for the bistable system, particularly for moderate volumes, e.g., $V$ in the range $0.5-30$. We note that simulating systems with very large or very small volumes is still a challenging task for variational quantum algorithms due to the limited precision available on current NISQ hardware---variational quantum algorithms must be able to distinguish between the eigenvalues of orthogonal states to reliably compute the desired eigenvalues and eigenvectors of the system.

\subsection{Eigenvalues}
\begin{figure*}[ht]
    \centering
    \subfloat[Original, non-Hermitian Schl\"ogl Operator.]{\includegraphics[width=0.45\textwidth]{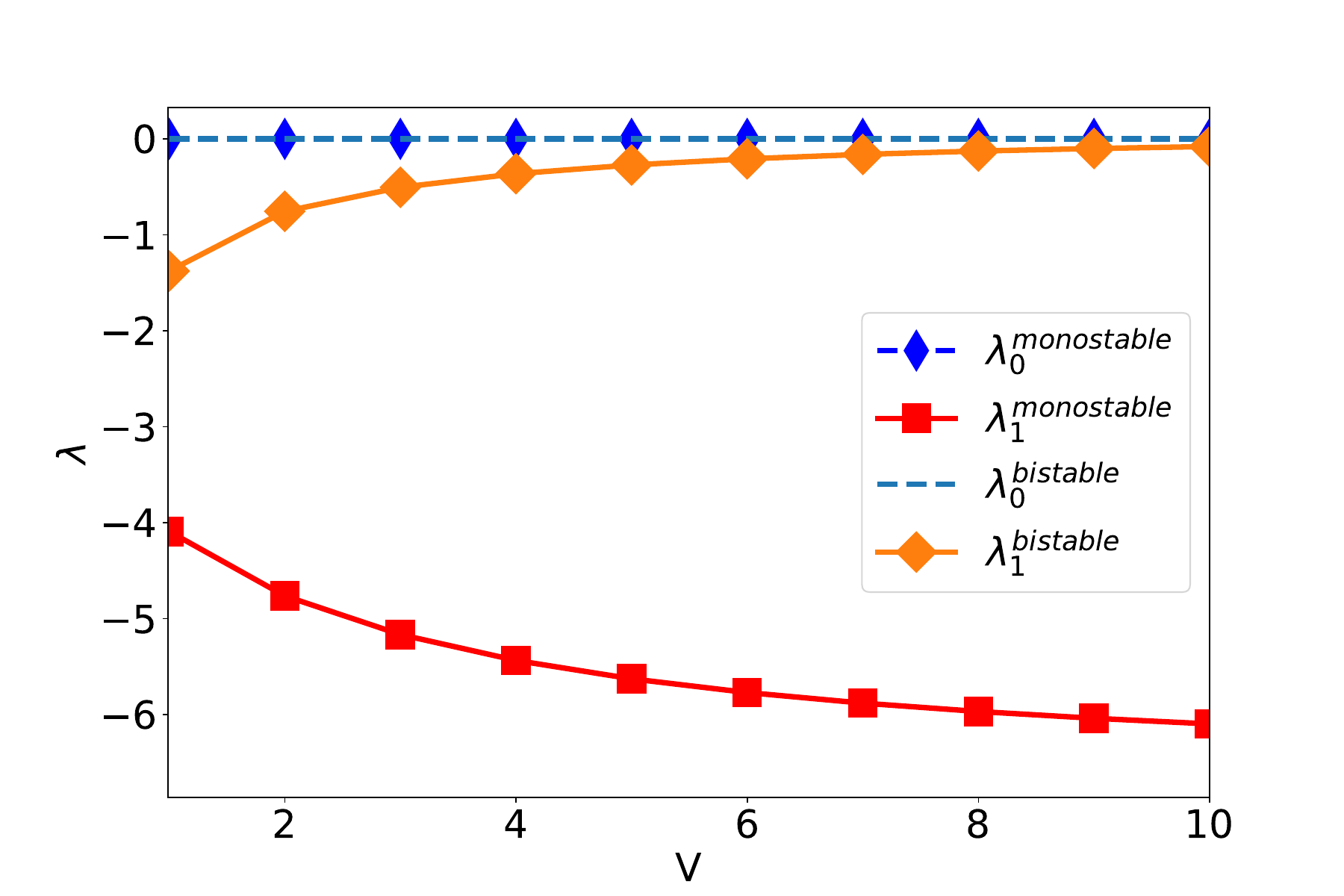}\label{fig:original-eigenvalues}}
    %\qquad
    \subfloat[Hermitian versus non-Hermitian eigenvalues.\label{fig:eig-comparison}]{\includegraphics[width=0.60\textwidth]{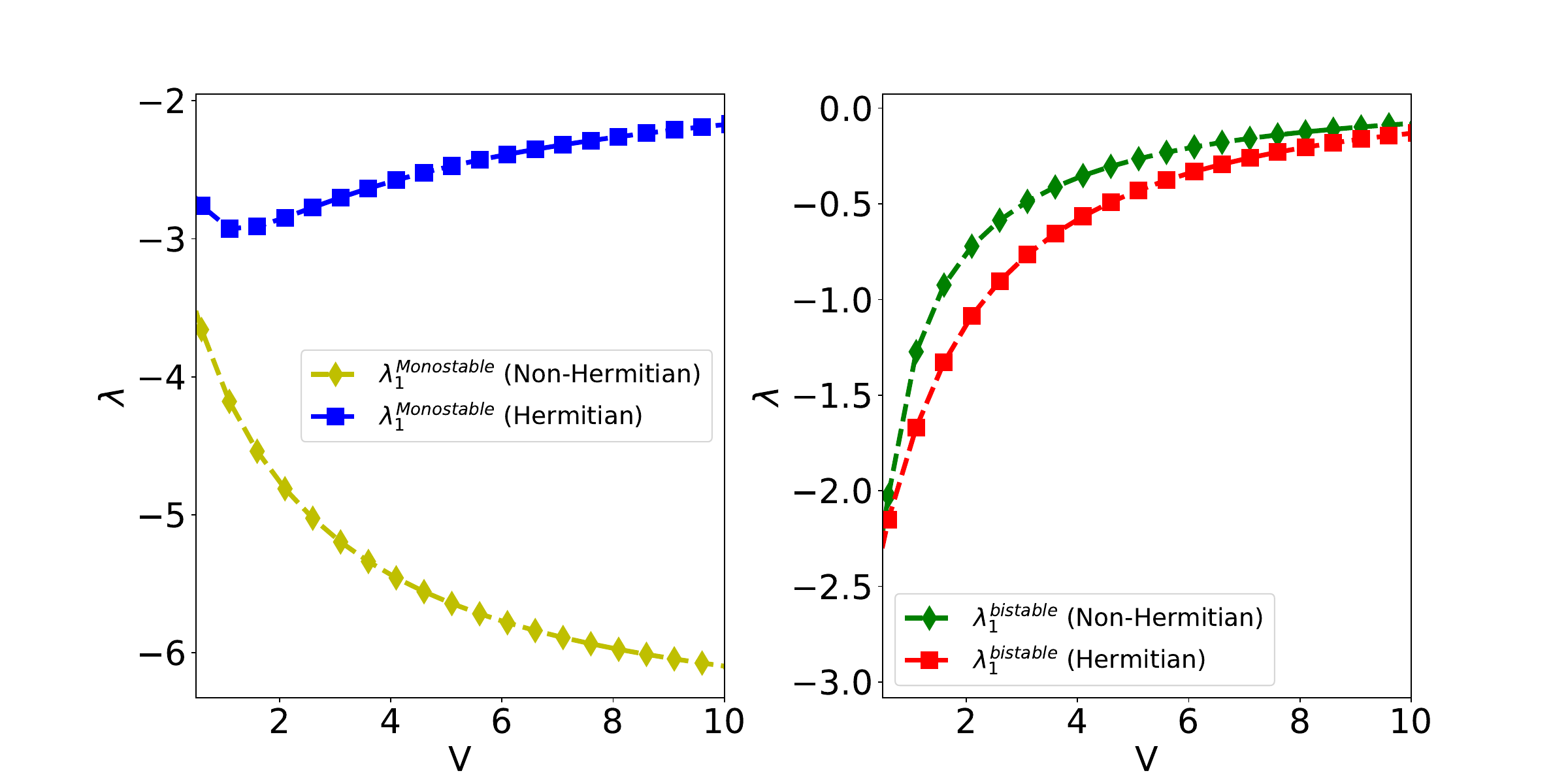}}
    \caption{Classically-computed eigenvalues of the Schl\"ogl operator for different volumes. (a) $\lambda_0$ and $\lambda_1$ from the original, non-Hermitian Schl\"ogl operator. (b) A comparison of the Hermitian and non-Hermitian values of $\lambda_1$. The gap between the monostable solutions is relatively small when $V$ is tiny, but continues to grow steadily as $V$ increases. In the bistable case, the Hermitian solution approaches the solution to the original non-Hermitian Schl\"ogl problem in the very large or very small volume limits. However, for large volumes, e.g., $V \gg 10$, numerical instabilities arise and quickly make the problem intractable. Similarly, setting $V=0$ is not only numerically impractical but also devoid of any physical meaning.}
    \label{fig:classical-eigenvalues}
\end{figure*}

%% Classical eigenvalues
\paragraph{Classical Eigenvalues} Figure \ref{fig:classical-eigenvalues} shows the eigenvalues computed using exact diagonalization of the Hermitian and non-Hermitian Schl\"ogl operators. In the bistable case, $\lambda_0$ and $\lambda_1$ become indistinguishable for very small or very large volumes.

\begin{figure*}
    \subfloat[2-qubit monostable operator.]{{\includegraphics[width=0.35\textwidth]{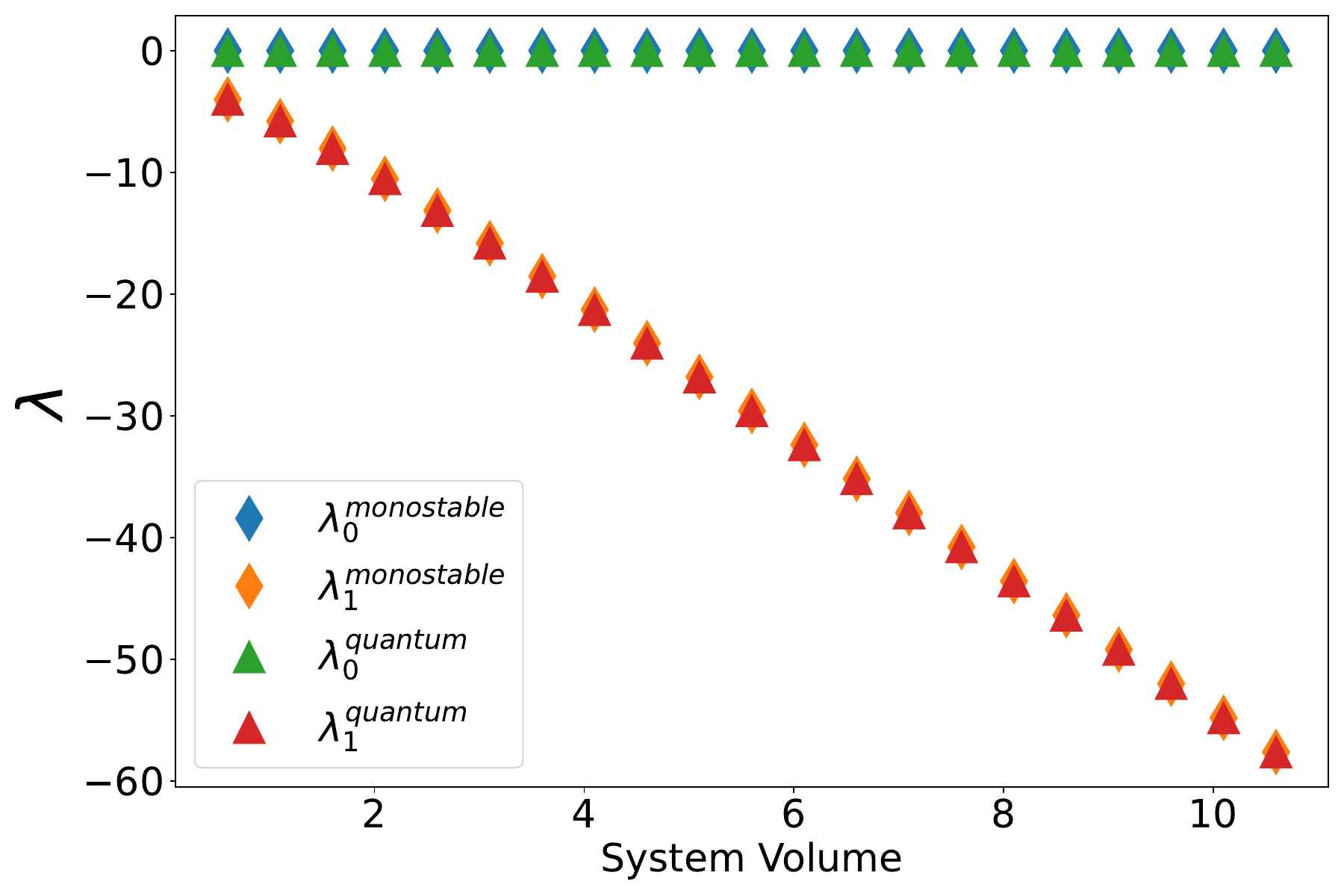} }}%
     \subfloat[2-qubit bistable operator.]{{\includegraphics[width=0.35\textwidth]{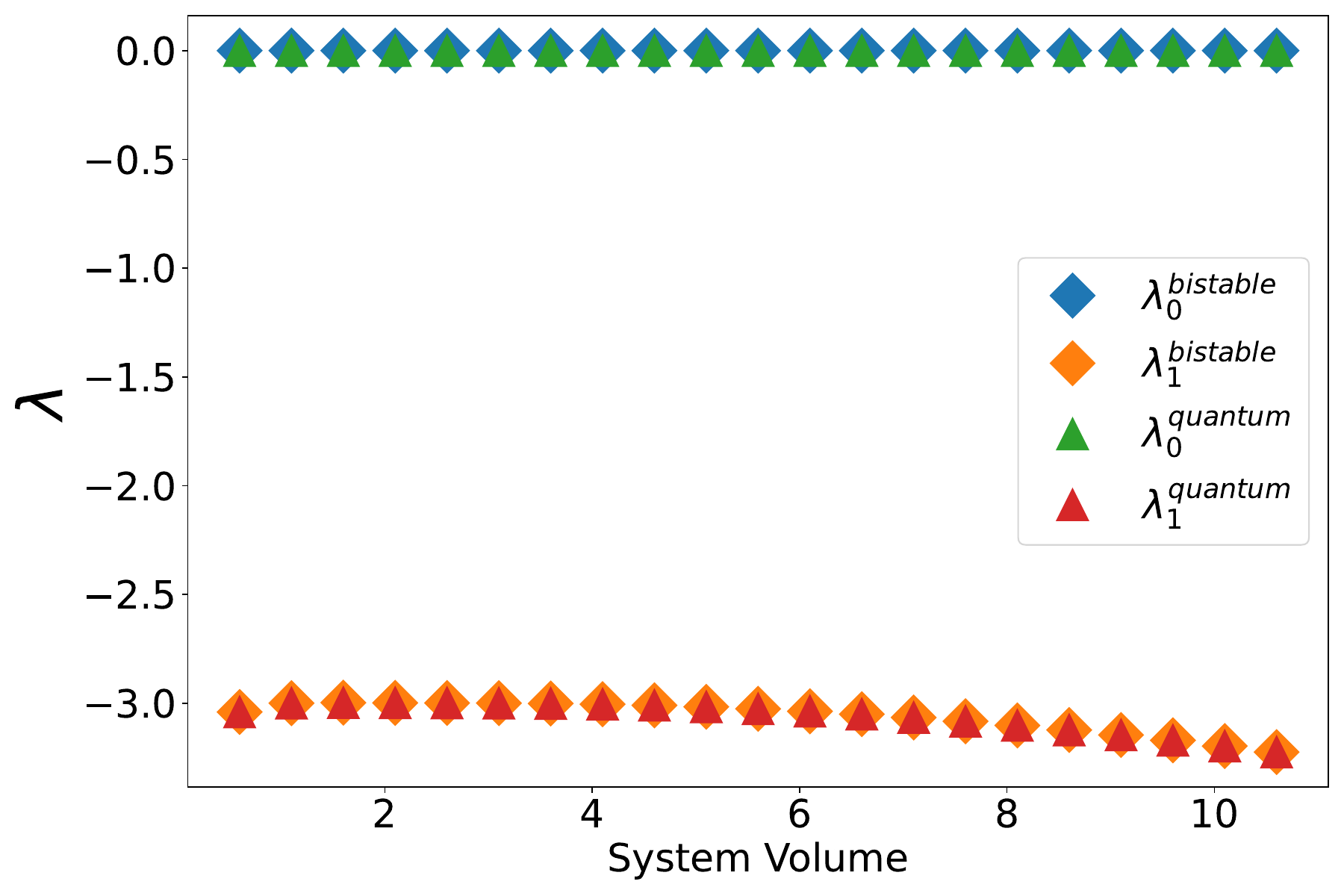} }}%
     \hfill
    \subfloat[3-qubit monostable operator.]{{\includegraphics[width=0.35\textwidth]{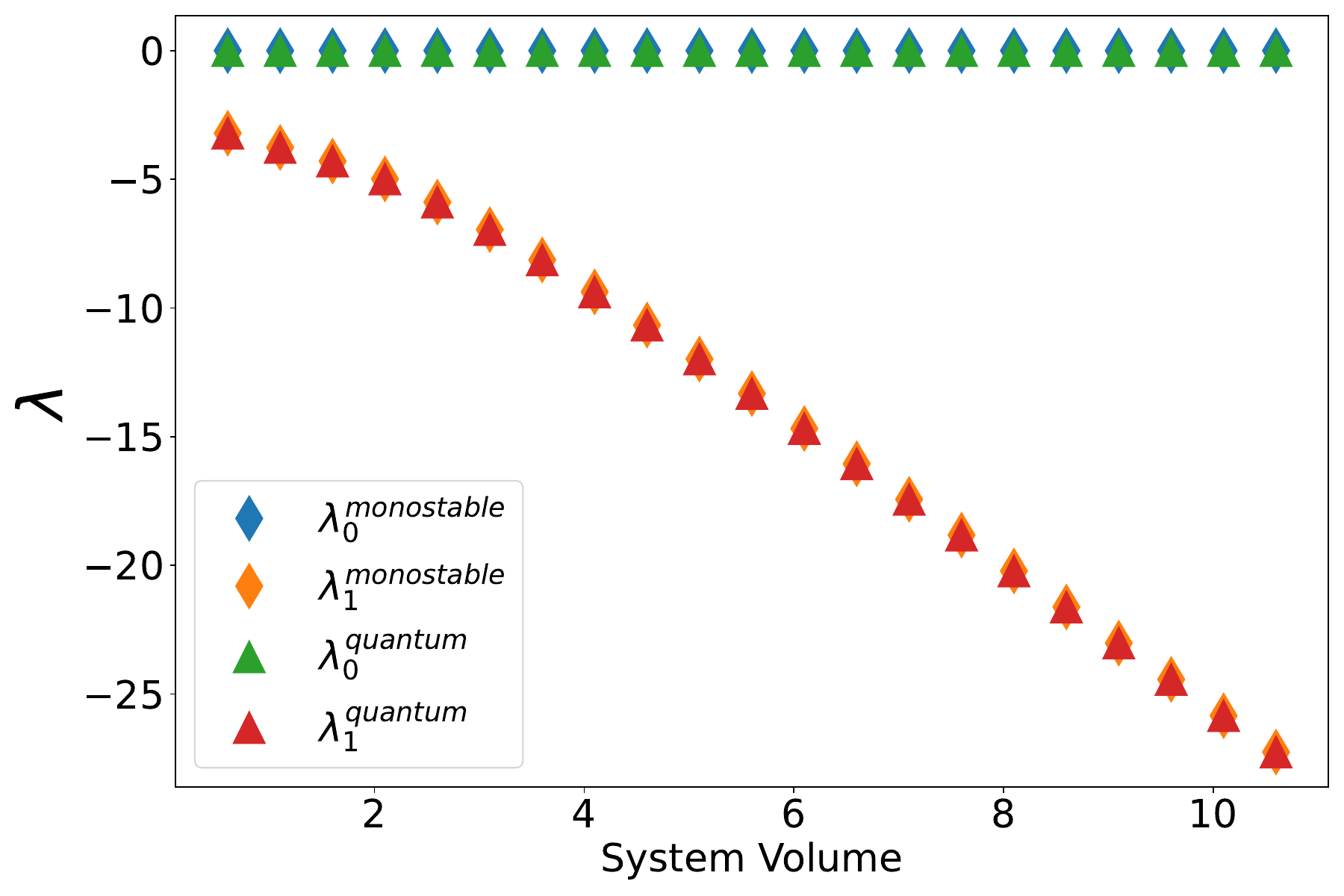} }}%
    \subfloat[3-qubit bistable operator.]{{\includegraphics[width=0.35\textwidth]{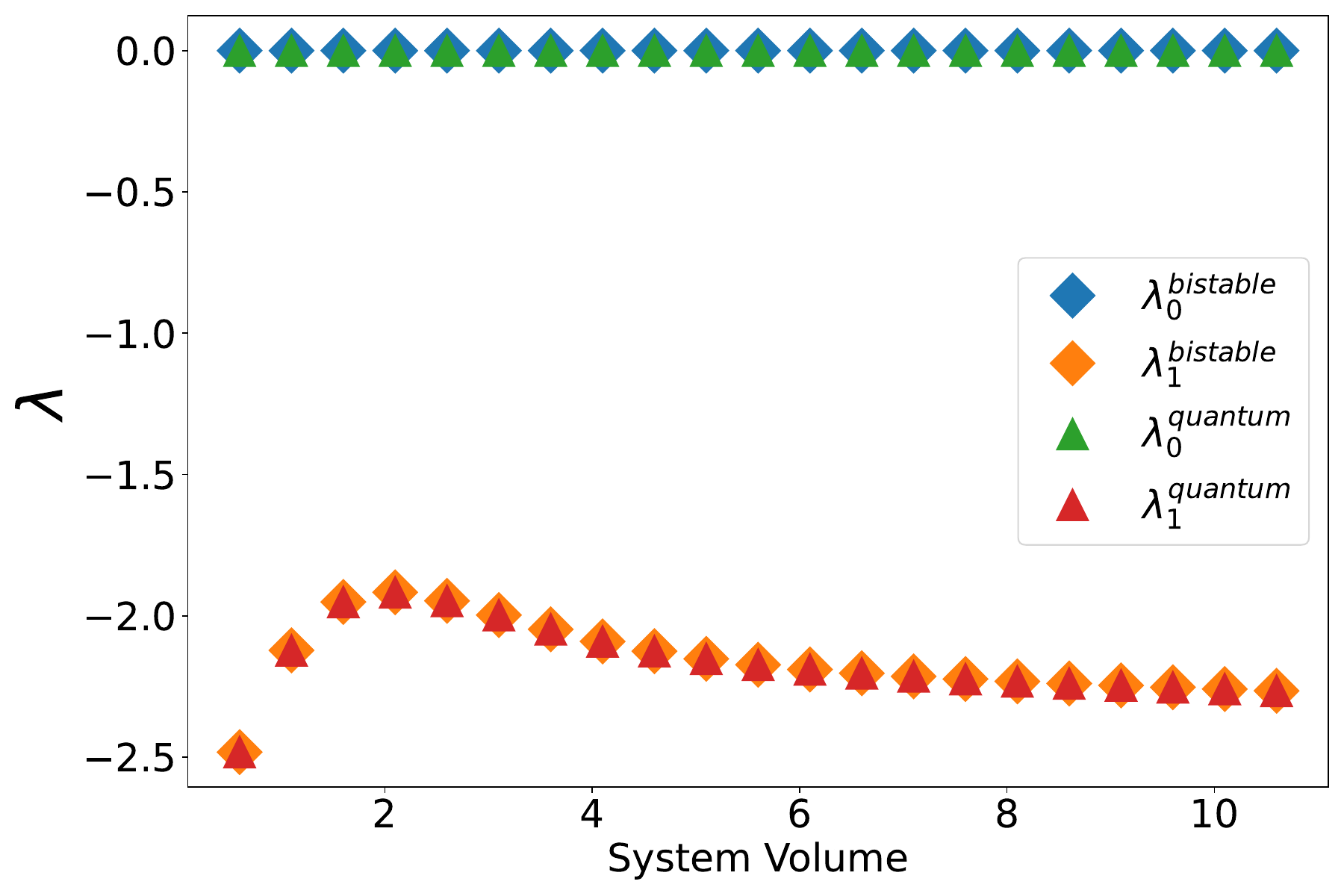} }}%
    \hfill
    \subfloat[4-qubit monostable operator using ans\"atz A.]{{\includegraphics[width=0.35\textwidth]{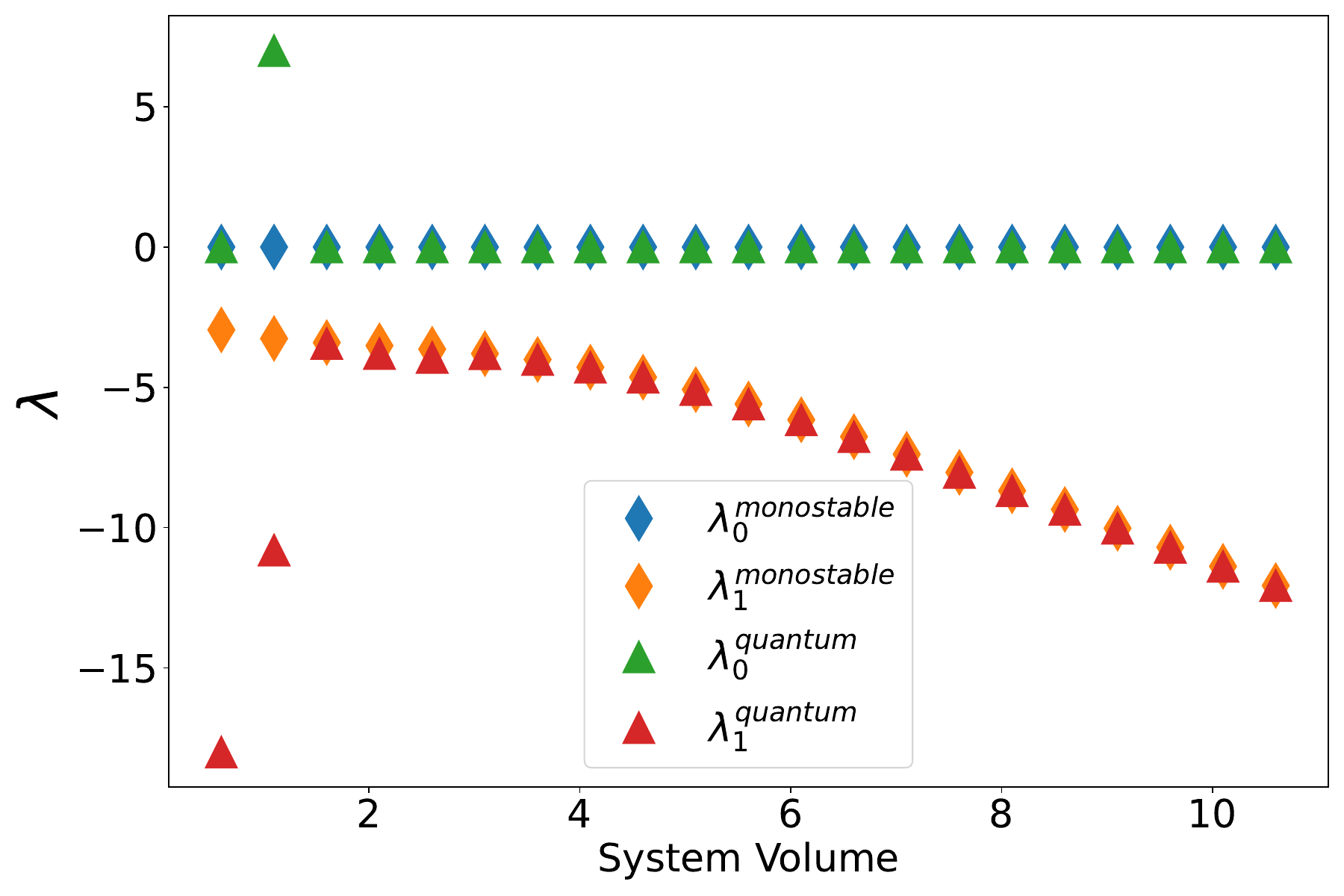} }}%
    \subfloat[4-qubit bistable operator using ans\"atz A.]{{\includegraphics[width=0.35\textwidth]{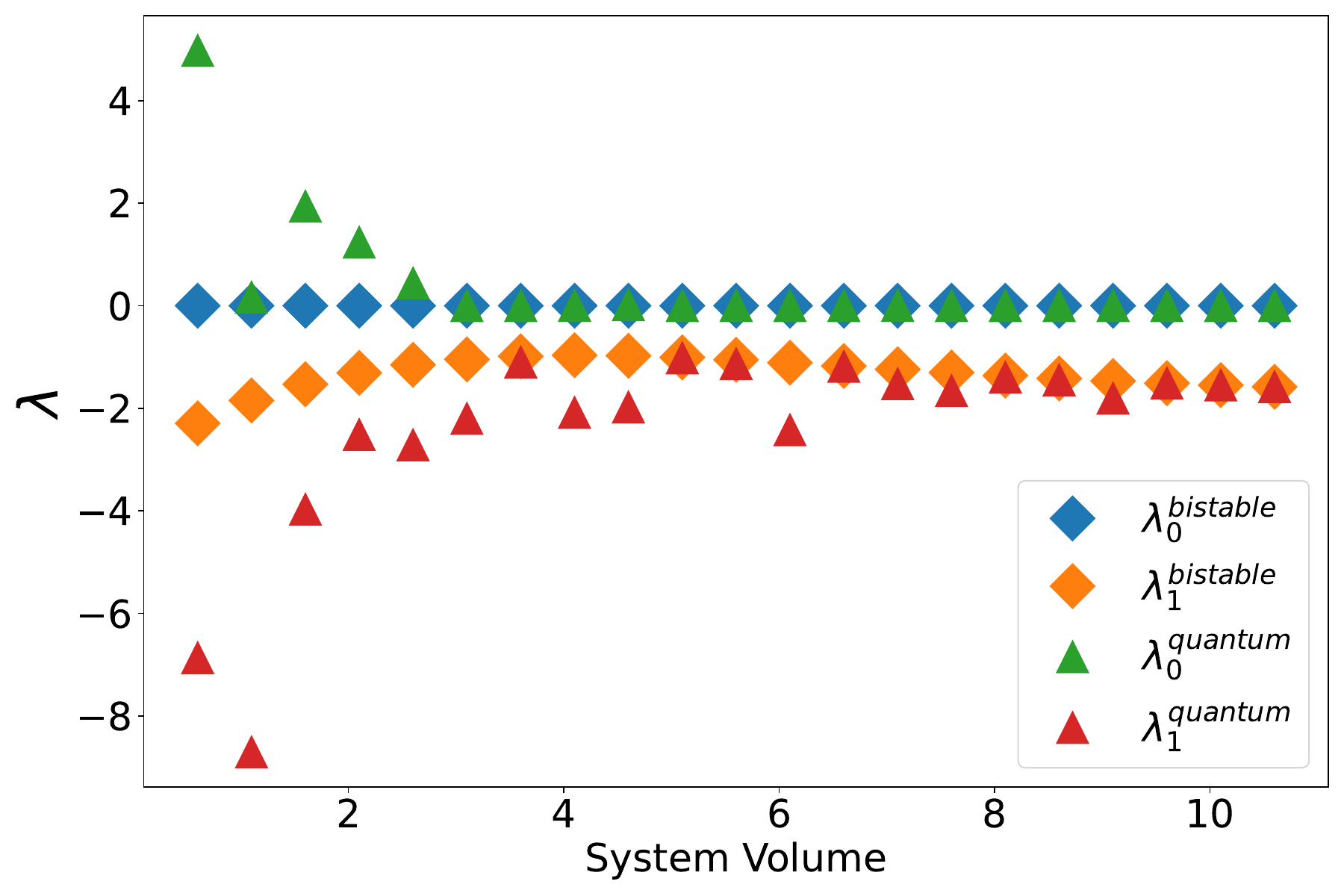} }}%
    \hfill
    \subfloat[4-qubit monostable operator using ans\"atz B.]{{\includegraphics[width=0.35\textwidth]{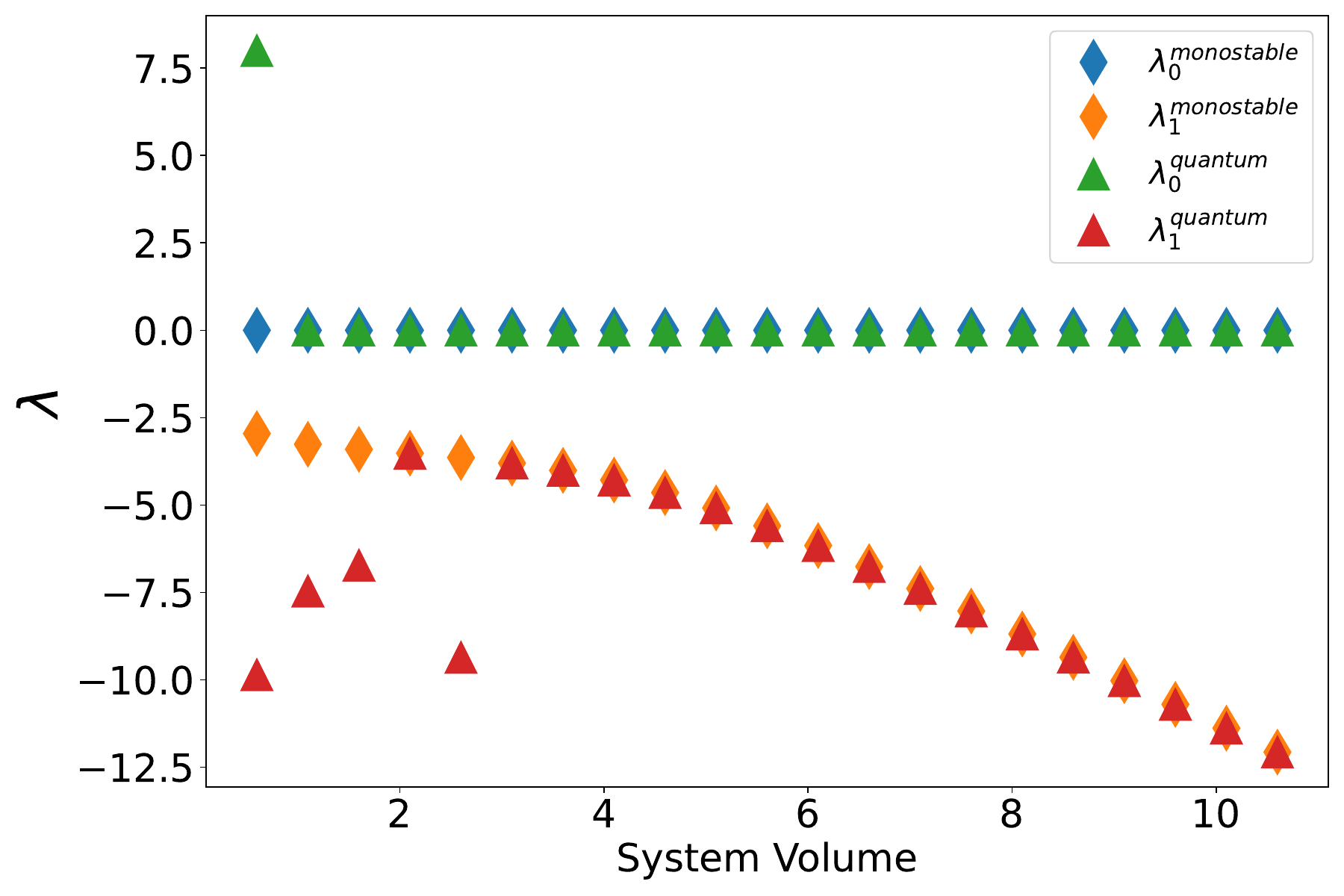} }}%
    \subfloat[4-qubit bistable operator using ans\"atz B.]{{\includegraphics[width=0.35\textwidth]{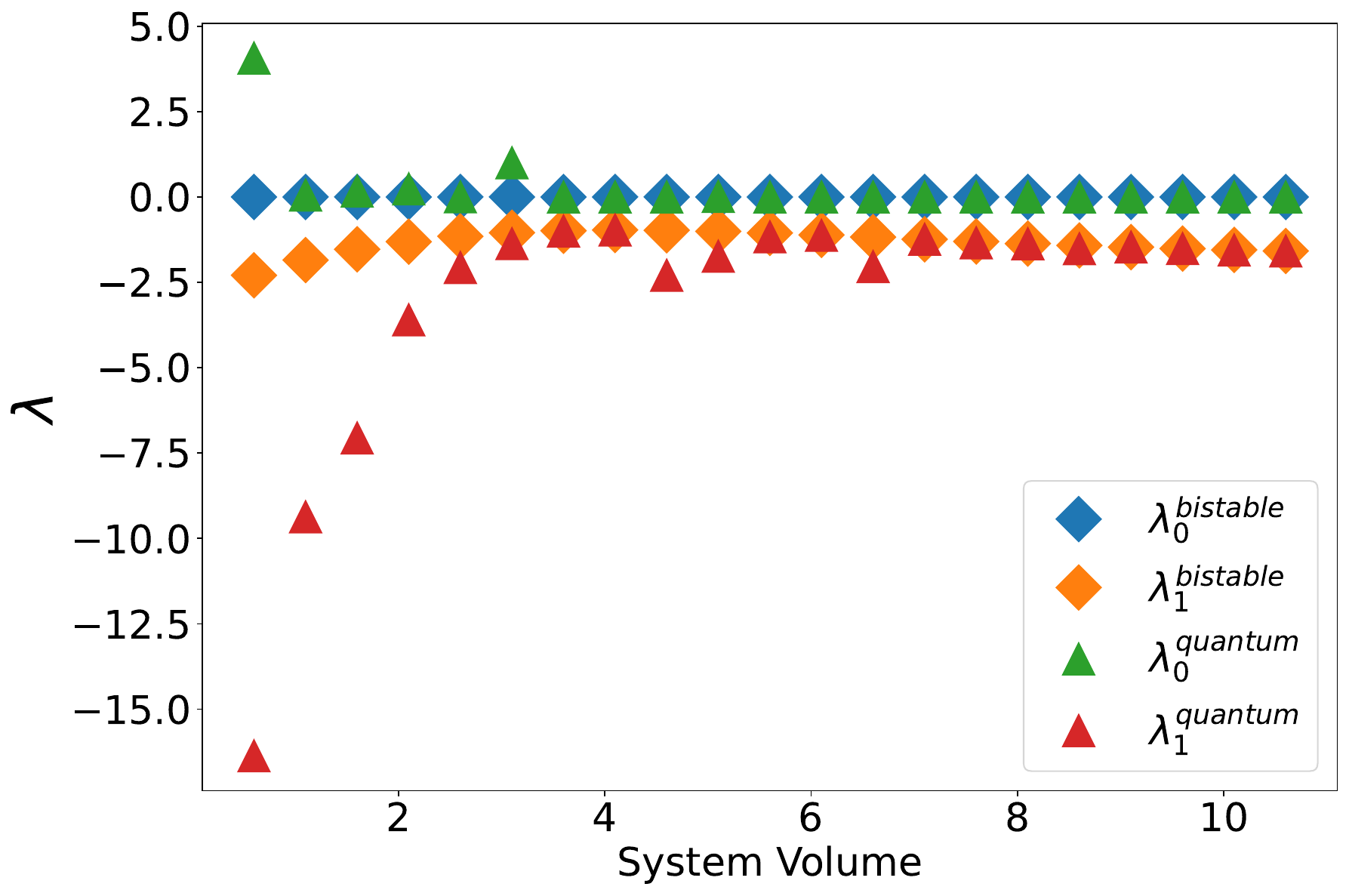} }}%
    \caption{Quantum versus classical eigenvalues for the monostable and bistable Schl\"ogl operators using 2- ((a) and (b)), 3- ((c) and (d)), and 4- ((e)-(h)) qubit basis sets. The eigenvalues computed using VQD are in excellent agreement with the classically-computed eigenvalues for the 2- and 3-qubit cases. Numerically exact results are obtained for these operators for all system volumes considered. For the 4-qubit operators, two different ans\"atze sizes, A and B, were used corresponding to the {\tt TwoLocal} ans\"atz types with 4 and 5 circuit repetitions, respectively. The low accuracy observed for small volumes is likely due to the limited expressivity of the ans\"atz. Ans\"atz A performed slightly better than B for these volumes but ans\"atz B was more stable when the system volume was moderately larger. Details about our ans\"atz analysis for different operators can be found in the Supplementary Materials.}
    \label{fig:quantum-eigenvalues}
\end{figure*}

% quantum eigenvalues
\paragraph{Quantum Eigenvalues}
The first two lowest eigenvalues for the Hermitian bistable Schl\"ogl model were computed using VQD. The 2-, 3-, and 4-qubit operators were used to approximate the Schl\"ogl operator as shown in Figure \ref{fig:quantum-eigenvalues}. These operators were first converted to Hermitian form using Equation \ref{eq:Qspd} and then decomposed into Pauli strings. The quantum computed eigenvalues are in excellent agreement with the classical (Hermitian) eigenvalues for the 2- and 3-qubit operators. The errors observed in the 4-qubit operator case can be attributed to a combination of a poor ans\"atz and the computational difficulty of optimizing variational parameters for a large circuit -- both well-known issues in the literature.\cite{fedorov_vqe_2022, tilly_variational_2022, cao_quantum_2019} For context, the 2-qubit ans\"atz had only 4 variational parameter to optimize, while the 4-qubit ans\"atz had up to 24 parameters.

For the 2- and 3-qubit operators, various sorting techniques were adopted to analyze which terms can be omitted from the operator in order to boost our computational efficiency without significantly sacrificing accuracy. In the positive-first sort, terms were arranged from largest to smallest, starting with positive values. In the default sort, no modifications were made to the order of the terms after applying the Pauli decomposition algorithm in Qiskit. In the magnitude sort, the Pauli terms were arranged in order of the magnitudes of the coefficients of the Pauli terms. In the optimized sort, only terms that give non-zero expectation values with respect to the known initial state given in Equation \ref{eq:initial_state} were used. Since optimized sort requires classically computing the expectation values of each Pauli term, it was only used as an additional check to understand what the most important terms were for an exact evaluation of $\lambda_0$. Figure \ref{fig:error_plots} shows the errors on the associated quantum computed eigenvalues. Table S4 in the Supplementary Materials shows the lowest percent errors obtained for each sorting technique. For $\lambda_0$, default sorting the Pauli terms and using only $\sim 50\%$ of them was sufficient to obtain an exact solution. We empirically found that the expectation values of at least the latter half of any operator were zero and, thus, did not contribute to $\lambda_0$. This was not the case for $\lambda_1$ since all the expectation values were non-zero. Nonetheless, we found that for $\lambda_1$, sorting the operators by the magnitude of their coefficients gave the most favorable trade-off between accuracy and the number of terms truncated from the original operator. As shown in Table S4 in the Supplementary Materials, using $24/28$ terms resulted in an error of $1.29\%$ for the 3-qubit operator and using $7/10$ terms resulted in a $0\%$ error in $\lambda_1$ for the 2-qubit operator. 

\begin{figure*}
    \centering
    
    % VQD iterations
    \subfloat[2-qubit bistable operator.]{{\includegraphics[width=0.33\textwidth]{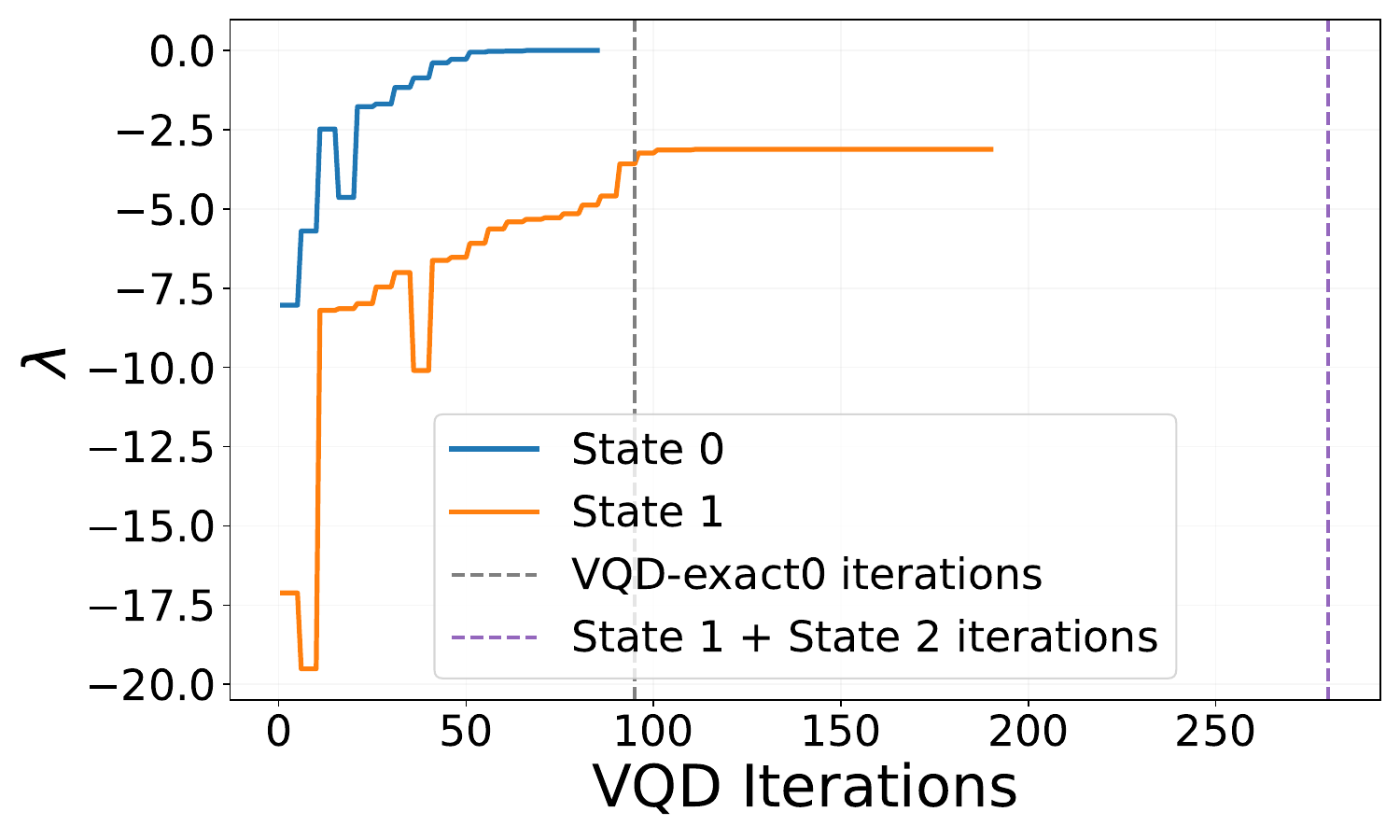} }}%
    \subfloat[3-qubit bistable operator.]{{\includegraphics[width=0.33\textwidth]{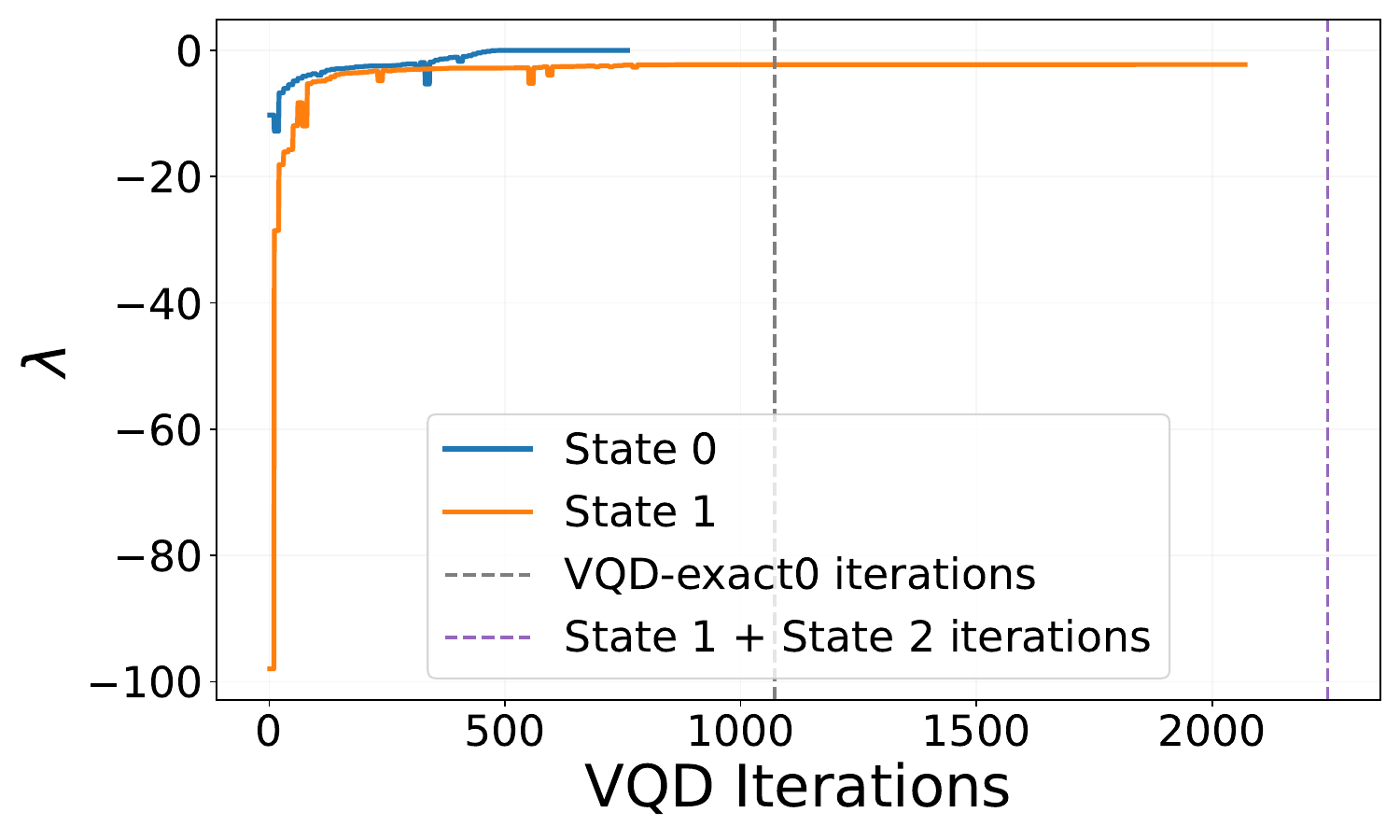} }}%
    \subfloat[4-qubit bistable operator.]{{\includegraphics[width=0.33\textwidth]{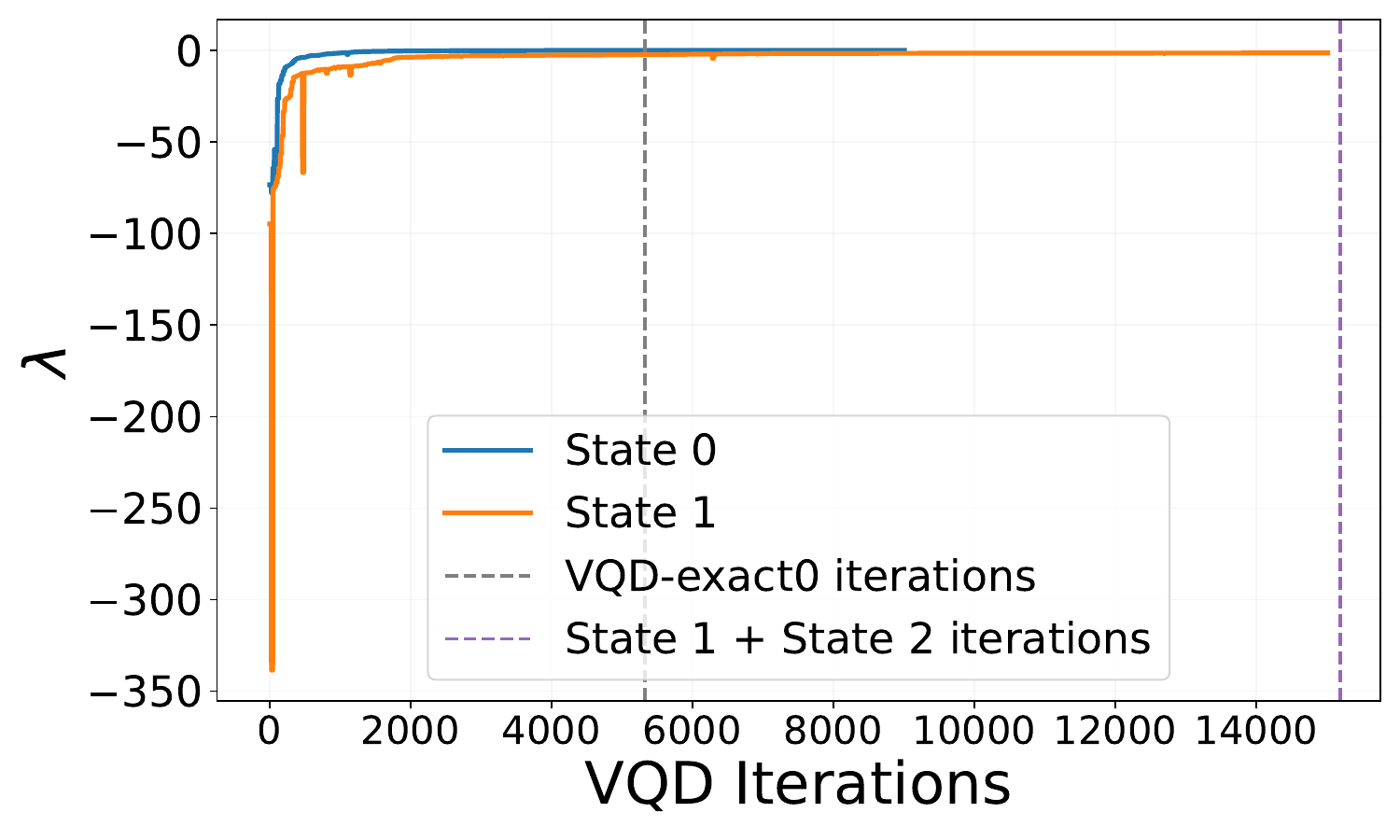} }}%
    
    % VDQ vectors for lambda1
    \subfloat[2-qubit operator.]{{\includegraphics[width=0.33\textwidth]{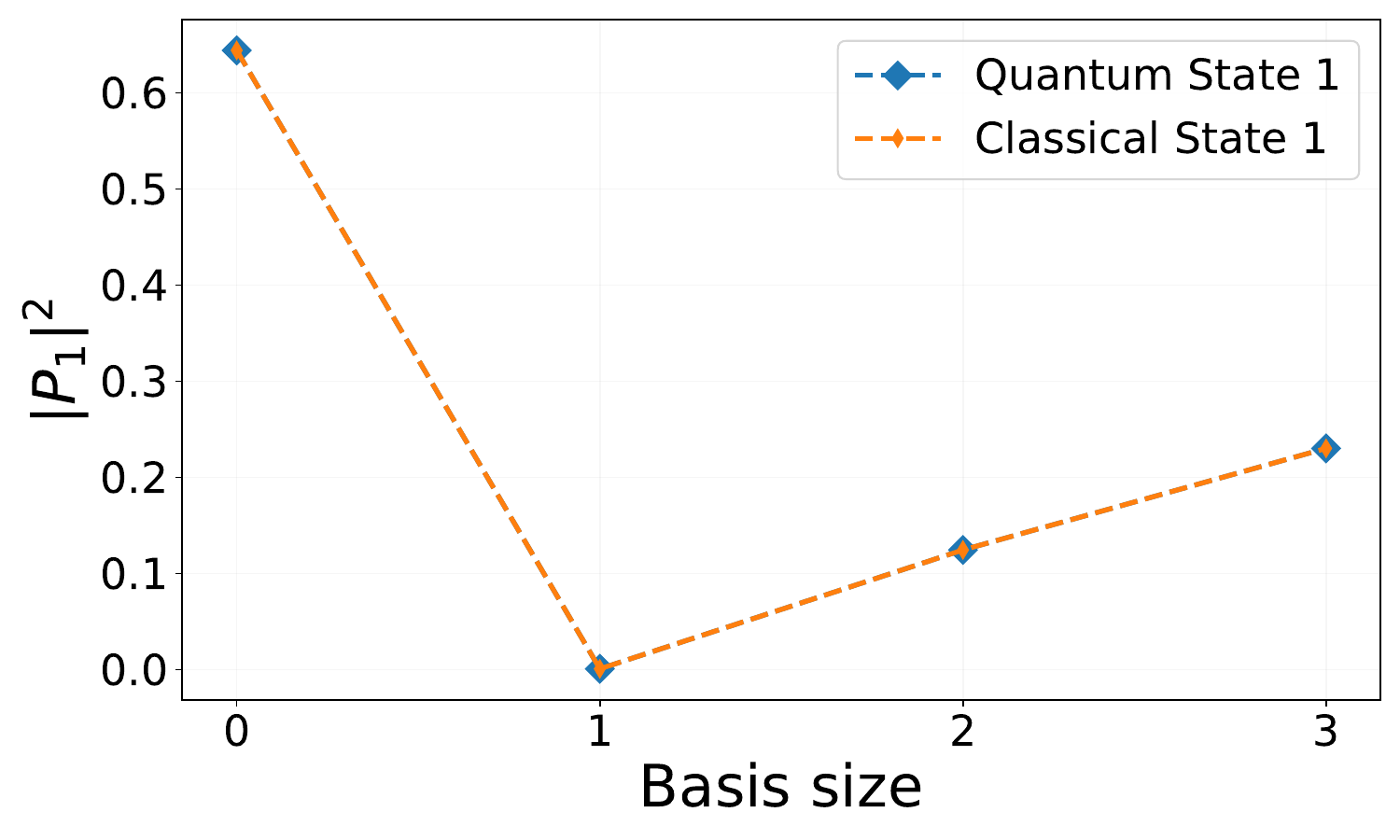} }}%
    \subfloat[3-qubit operator.]{{\includegraphics[width=0.33\textwidth]{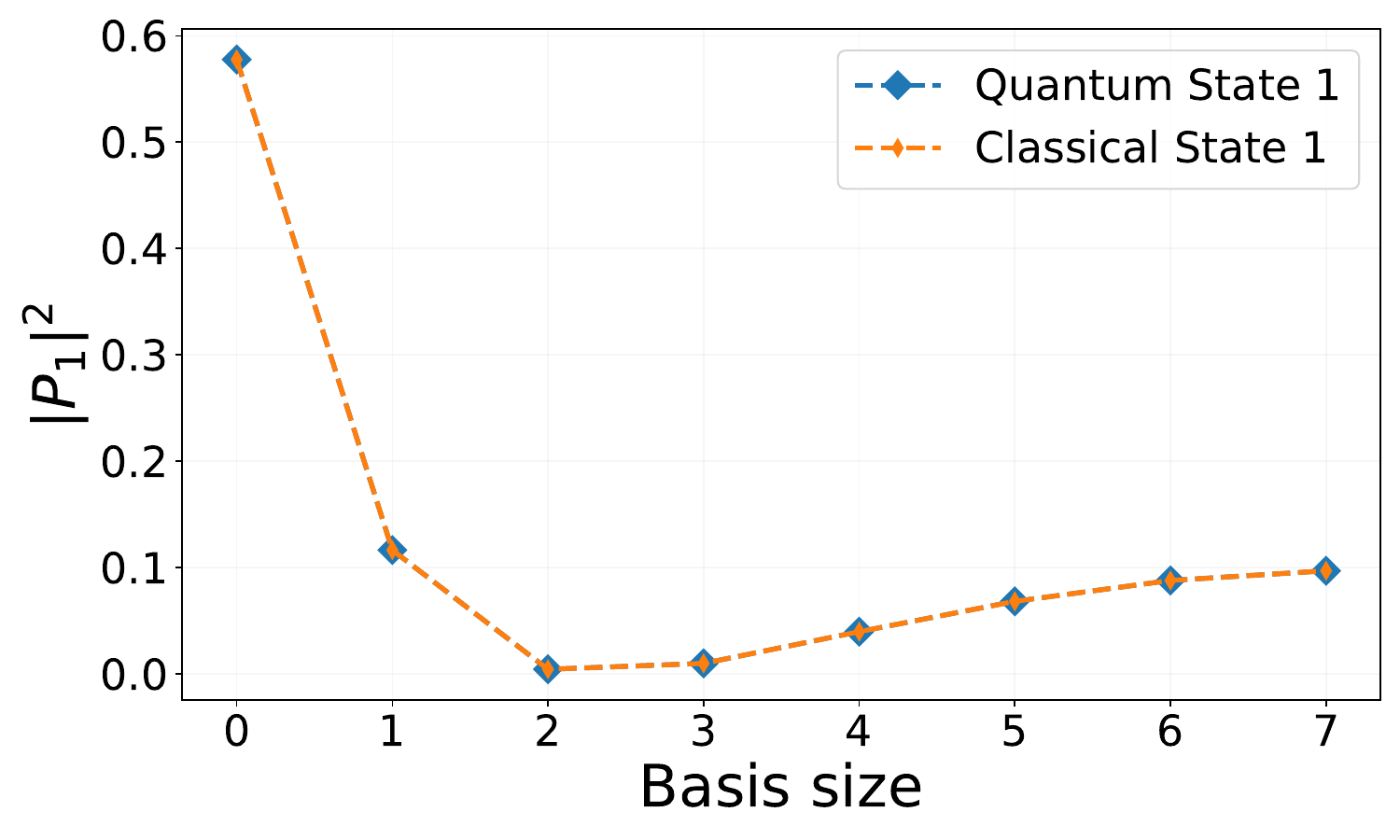} }}%
    \subfloat[4-qubit operator.]{{\includegraphics[width=0.33\textwidth]{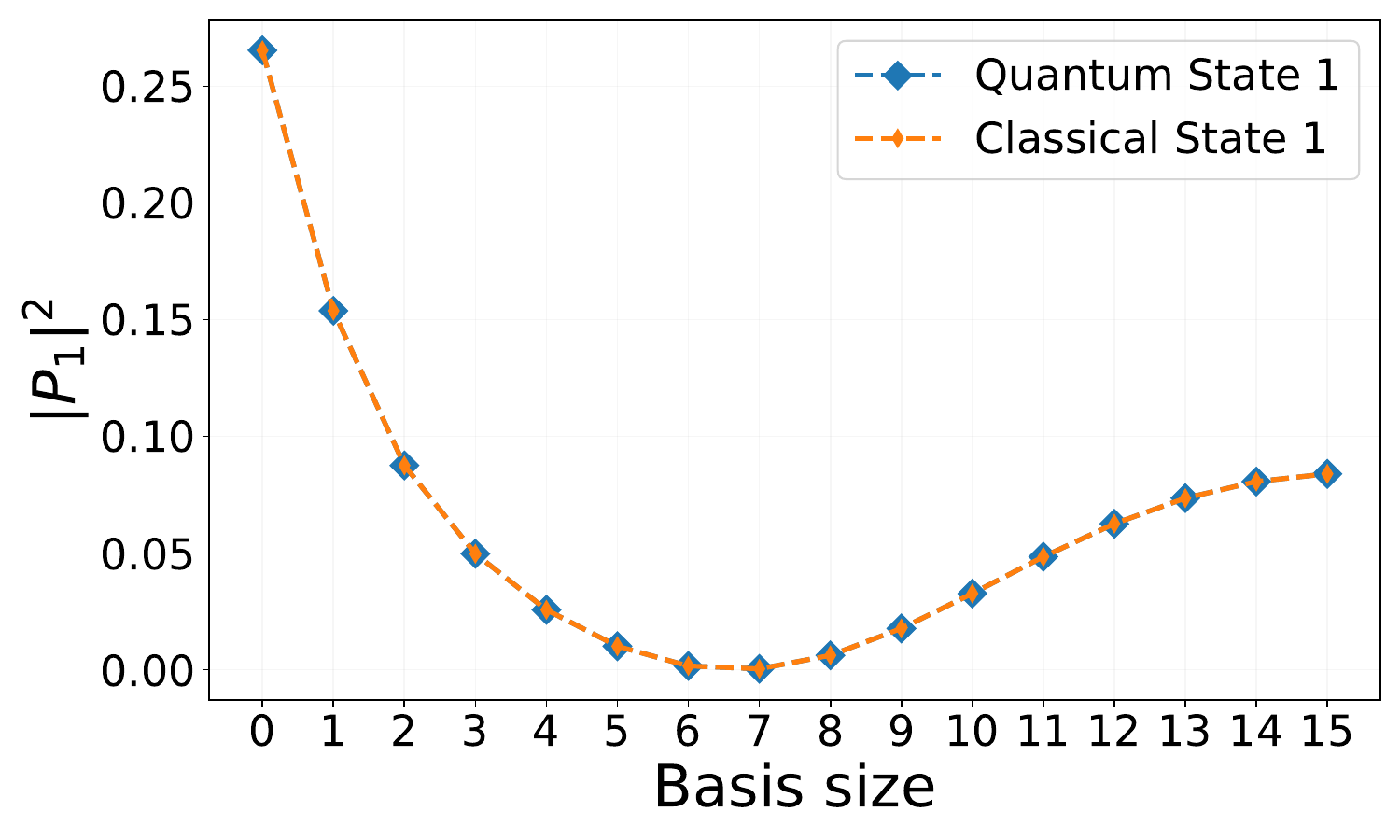} }}%

    \caption{Noiseless simulations with the VQD-{\tt exact0} algorithm. (a)-(c) show VQD iterations and (d)-(f) depict the eigenvectors corresponding to $\lambda_1$ for different basis sizes. The purple and black vertical lines represent the total number of iterations for the VQD and VQD-{\tt exact0} algorithms, respectively. Simulations for the 2-, 3-, and for 4-qubit operators with $V=8.5$ are shown. The number of VQD iterations increases rapidly with the basis size due to the increase in the number of variational parameters to be optimized. For VQD-{\tt exact0}, the iterations can be reduced significantly (by at least half) due to the exact initialization of the first excited state.}
    \label{fig:vqd_convergence}
\end{figure*}

\paragraph{VQD-{\tt exact0}} Finally, we computed the eigenvalues for all three operators using VQD with an exact initial state (we label this implementation as VQD-{\tt exact0}). We utilized our knowledge of the zeromode of the Hermitian operator given by Equation \ref{eq:Qspd} to accelerate the convergence of VQD. Specifically, we used Equation \ref{eq:initial_state} as the VQD initial ``guess" and observed a significant reduction in the number of VQD iterations. Figure \ref{fig:vqd_convergence} shows the convergence of $\lambda_0$ and $\lambda_1$ for 2-, 3-, and 4-qubit operators from our noiseless simulations. Exact values for $\lambda_0$ and $\lambda_1$ were obtained with VQD-{\tt exact0} in less than half the number of iterations needed by standard VQD. Additionally, highly accurate representations of the eigenvector corresponding to $\lambda_1$ were recovered from the circuit at the end of the optimization. Our numerical simulations showed an overall improvement in both the accuracy and convergence with VQD-{\tt exact0} even for the 4-qubit operator with the same ans\"atz\"e. Our findings highlight the importance of constructing an informed initial guess for the state vector to run and extract accurate results from variational quantum algorithms efficiently. 

%% Eigenvalue error plots
\begin{figure*}
   \subfloat[$\lambda_0$ from 2-qubit bistable operator.]{{\includegraphics[width=0.50\textwidth]{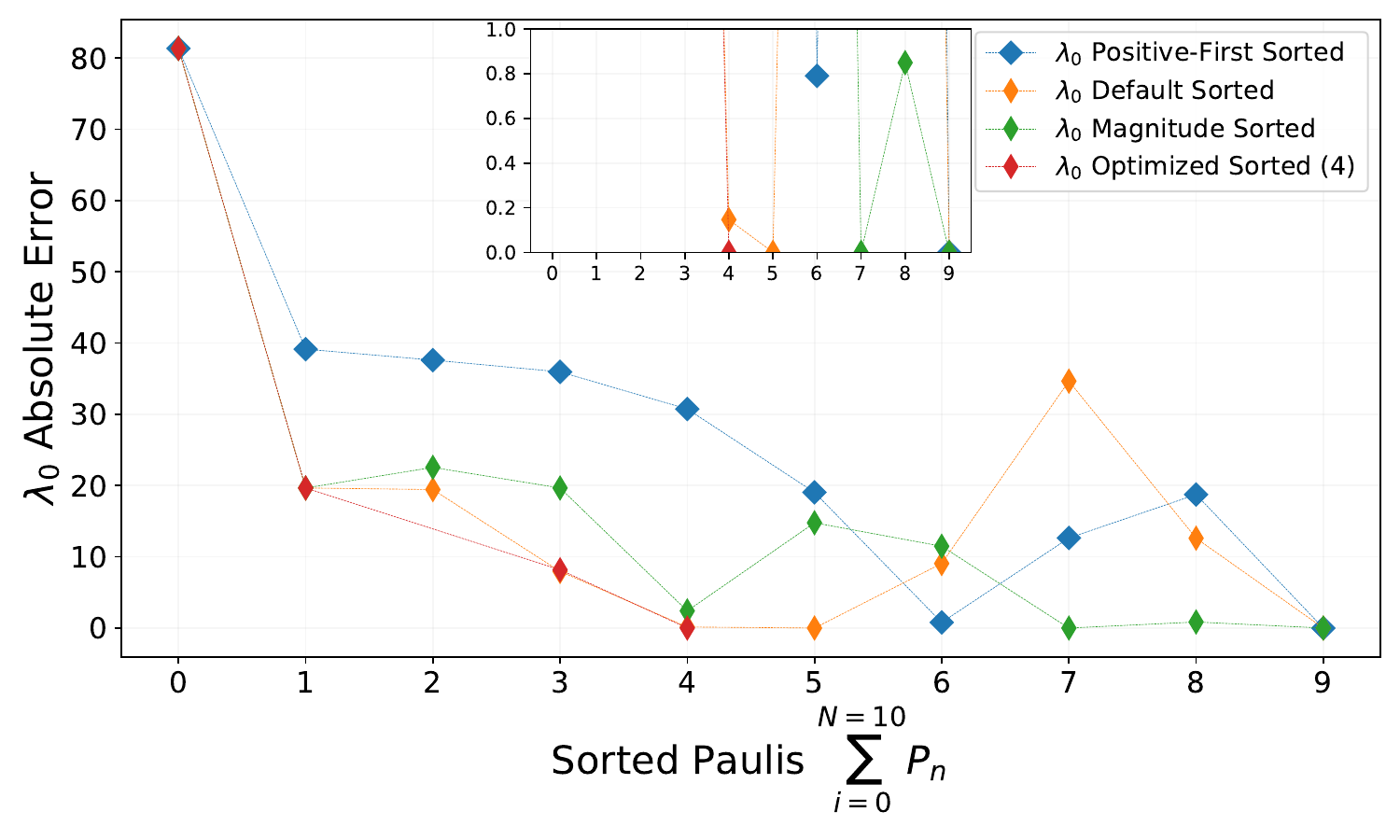} }}%
     \subfloat[$\lambda_1$ from 2-qubit bistable operator.]{{\includegraphics[width=0.50\textwidth]{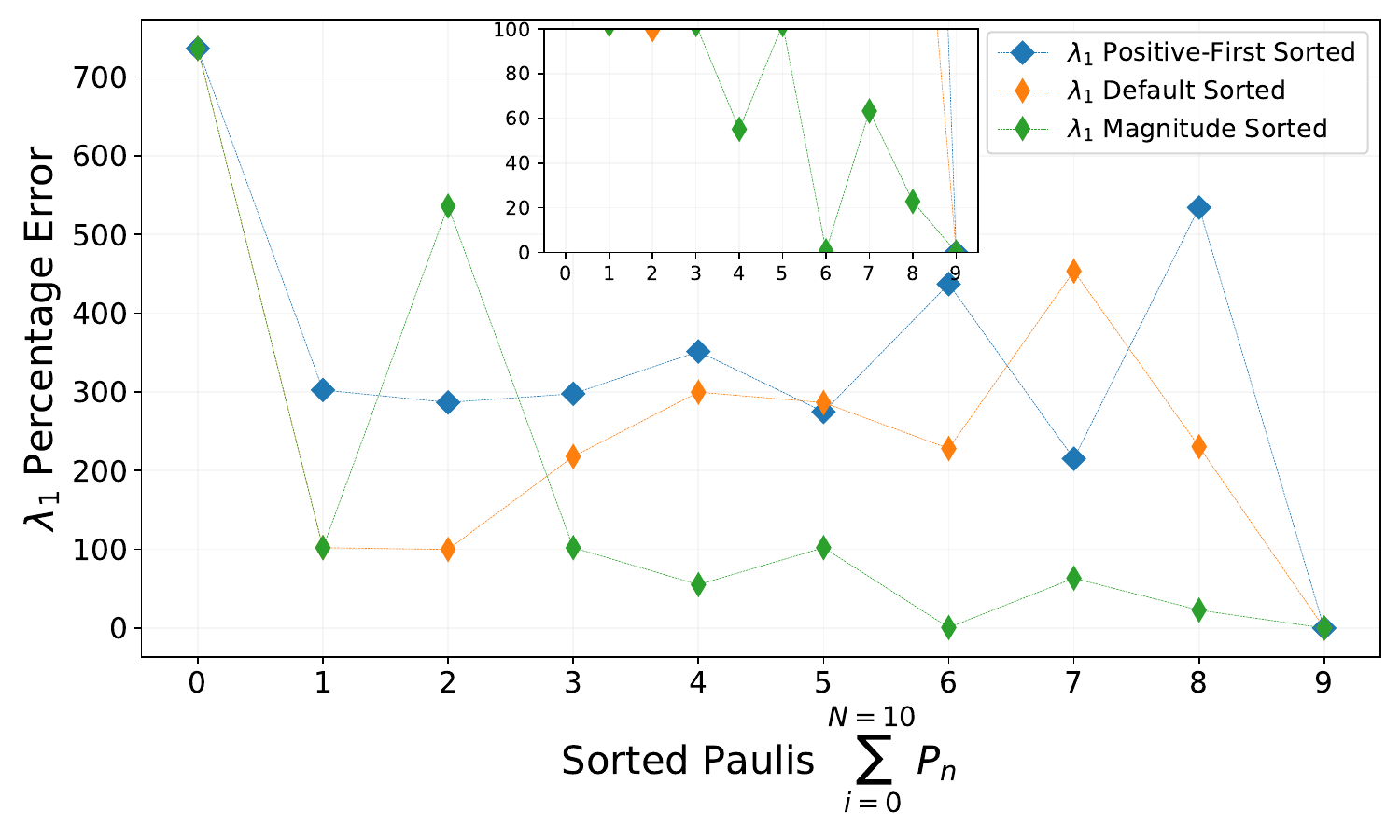} }}%
     \hfill
     \subfloat[$\lambda_0$ from 3-qubit bistable operator.]{{\includegraphics[width=0.50\textwidth]{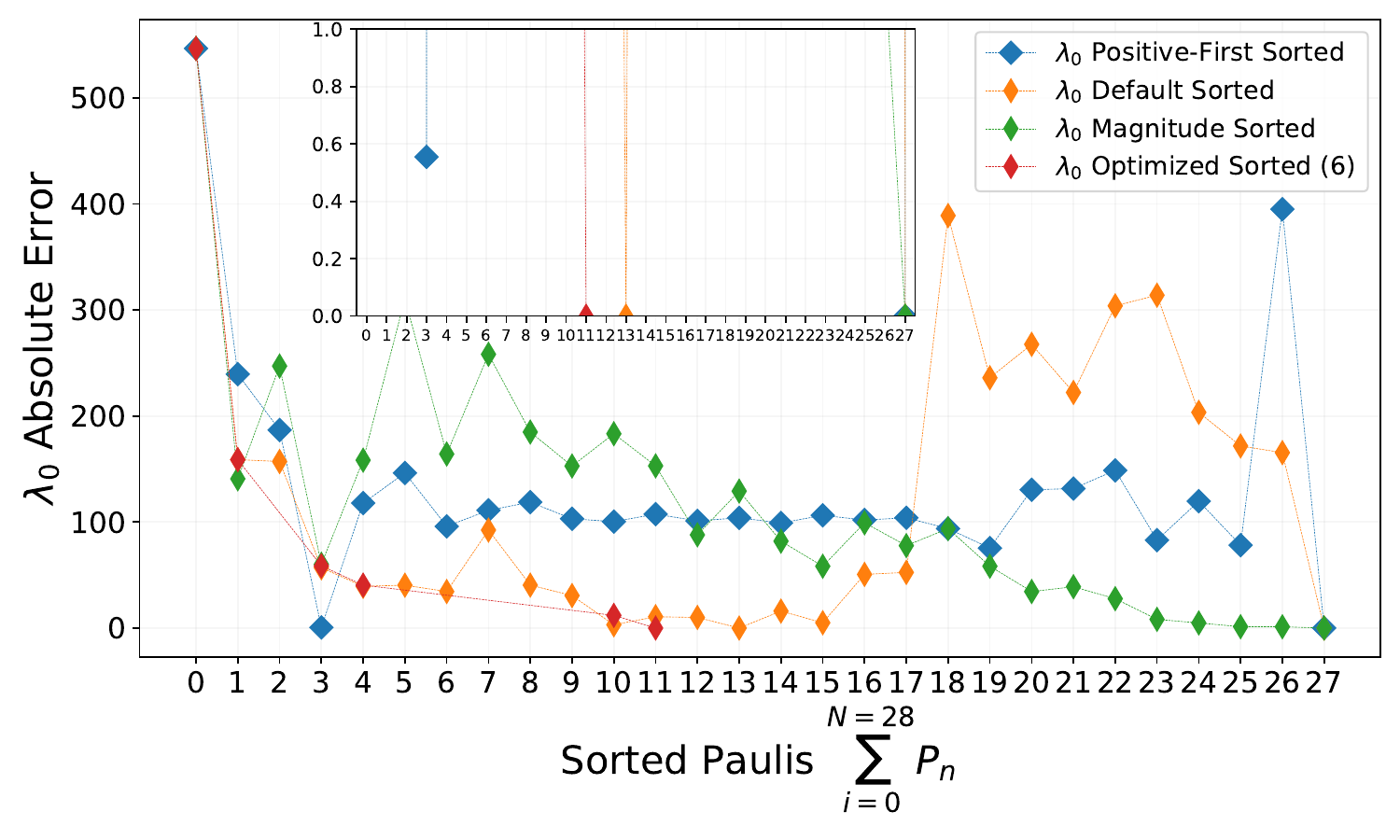} }}%
     \subfloat[$\lambda_1$ from 3-qubit bistable operator.]{{\includegraphics[width=0.50\textwidth]{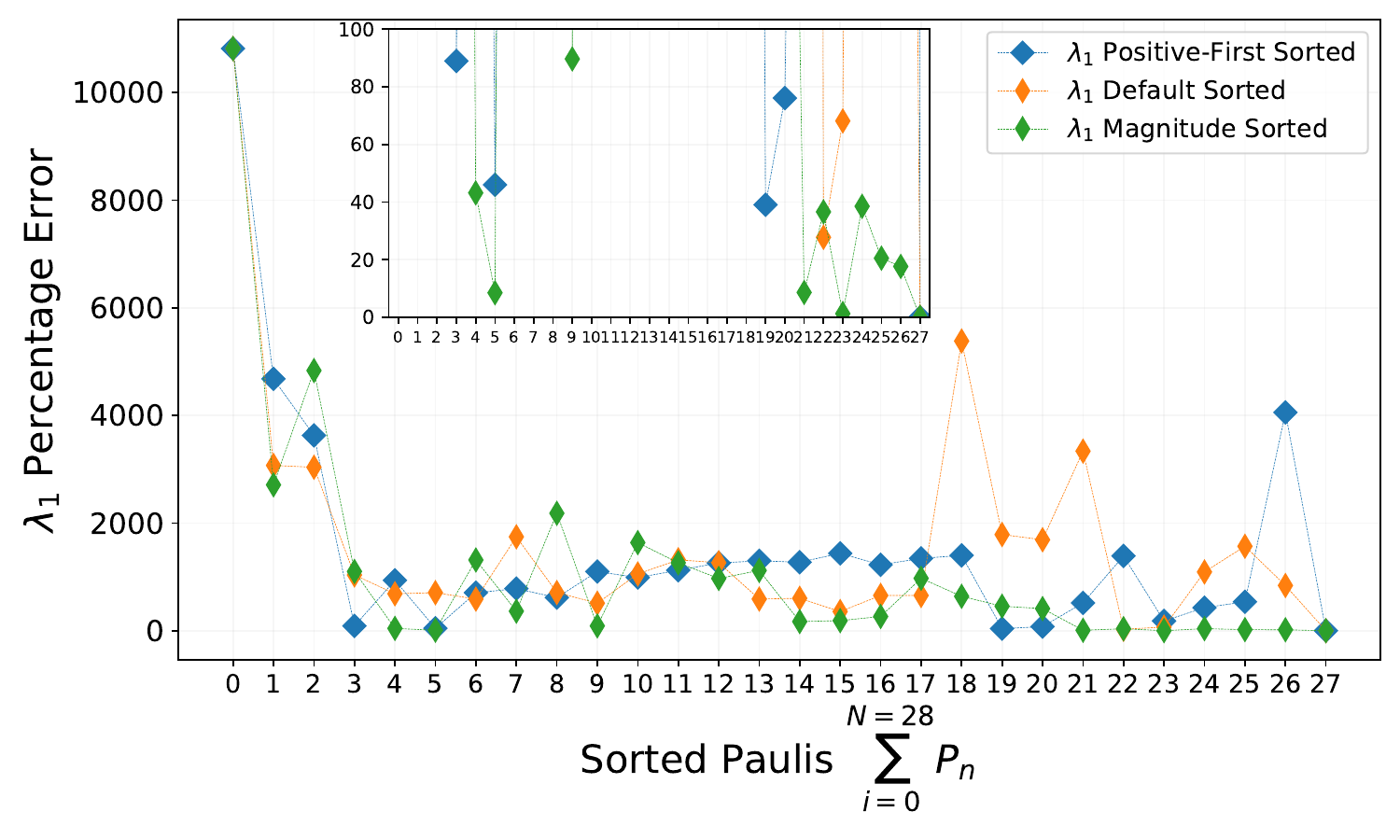} }}%
    \caption{Error plots for 2- and 3-qubit operators in $\lambda_0$ and $\lambda_1$ from different Pauli term sorting methods. (a) and (b) show the absolute errors in the estimation of $\lambda_0$ while (c) and (d) depict the percentage errors in the estimation of $\lambda_1$. The insets have the same units as the main plots. For $\lambda_0$, the default sort and optimized sort yield the most accurate results. In the default sort, only half of the terms are required for an exact solution. This is consistent for operators much larger than 3 qubits, and also holds true for the monostable operator. For $\lambda_1$, the magnitude sort provided the most reasonable trade-off between accuracy and number of terms used. See Table S4 in the Supplementary Materials.}
    \label{fig:error_plots}
\end{figure*}

%% VQD iterations and eigenvectors of lambda1

\subsection{Non-Equilibrium Steady State}
\label{Zeromode}
Here, we report numerical results for the zeromode (i.e., the non-equilibrium steady state) of the stochastic (bistable) Schl\"ogl matrix obtained on {\tt ibm\_brisbane}, a 127-qubit QPU. We use QPE to verify if the non-equilibrium steady-state of the stochastic Schl\"ogl model can be recovered on quantum hardware. It is possible to numerically estimate the zeromode (non-equilibrium steady-state) using QPE directly, although we realized that we could not do so accurately due to a failure of QPE to resolve small amplitudes and correctly recover the components of the desired eigenvector, especially in the presence of noise. This is somewhat anticipated given that QPE is expected to become advantageous on future, fault-tolerant hardware, not the near-term hardware used in this work. This prompted us to switch to VQSVD to extract the zeromode of the Schl\"ogl operator instead. We set the number of precision qubits used to construct the QPE circuit equal to seven. In what follows, we present zeromode results for three different system volumes, i.e., $V = 1.1$, $V = 5.5$, and $V = 10.5$, respectively. 

In each case, we computed the zeromode using a two-qubit operator, which resulted in a $4 \times 4$ stochastic Schl\"ogl matrix. Employing the transformation in Eq. \ref{eq:Qh} yields an $8 \times 8$ matrix, resulting in a basis size of eight. This gave us a query qubit register size of three for the QPE circuit. We compared the zeromodes obtained via exact diagonalization to the ones obtained via the implementation of QPE + VQSVD (details regarding the hyperparameters chosen to run VQSVD may be found in the Supplementary Materials). We construct the stochastic Schl\"ogl matrix from the CME using the procedure outlined in Sec. \ref{Matrix construction}. 

We would like to remark here that a basis size of eight is by no means sufficient to reproduce the full bistable dynamics of the Schl\"ogl model (i.e., the presence of two bistable states in the reaction network). A much larger basis size is needed to achieve this on both classical and quantum computers. To reproduce bistability, at least a 5-qubit representation of the original Schl\"ogl operator (i.e., Equation \ref{eq:schl3}) is needed. This results in a basis vector with at least 32 components. However, since Equation \ref{eq:schl3} must be brought to its Hermitian form as per Equation \ref{eq:Qh}, an extra qubit is necessary to accommodate a matrix twice as large. Consequently, a basis size of at least 64 (which can be encoded using 6 qubits) would be required to demonstrate bistability. Our goal here is to present results that follow as a proof-of-concept, illustrating the recovery of the non-equilibrium steady state of the Schl\"ogl model on current NISQ hardware, albeit with potentially added overheads and/or resource requirements.

\begin{figure*}[t]
    \centering
    \subfloat[$~V = 1.1$.]
    {{\includegraphics[width=0.33\textwidth]{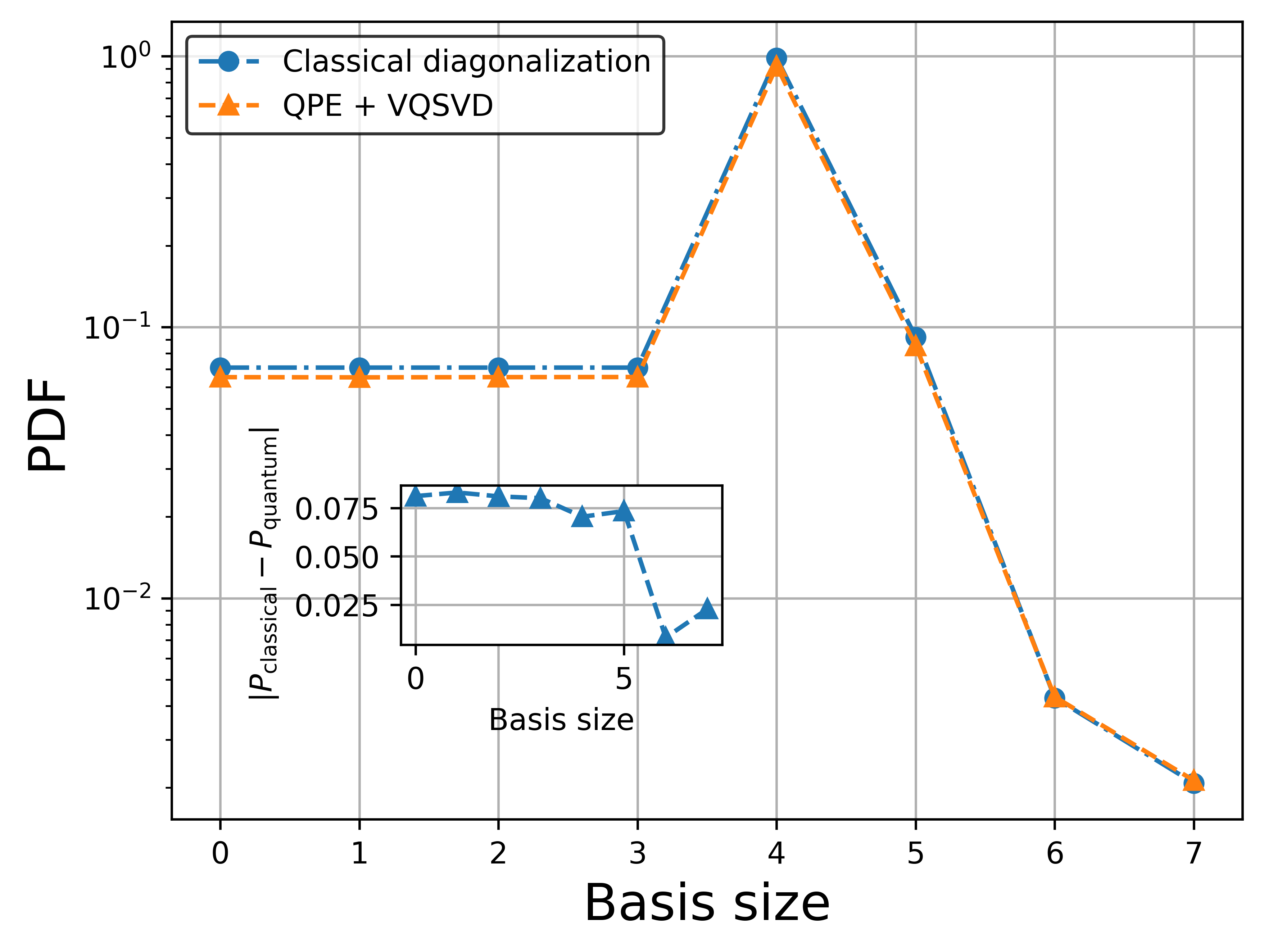} }}%
    \subfloat[$~V = 5.5$.]
    {{\includegraphics[width=0.33\textwidth]{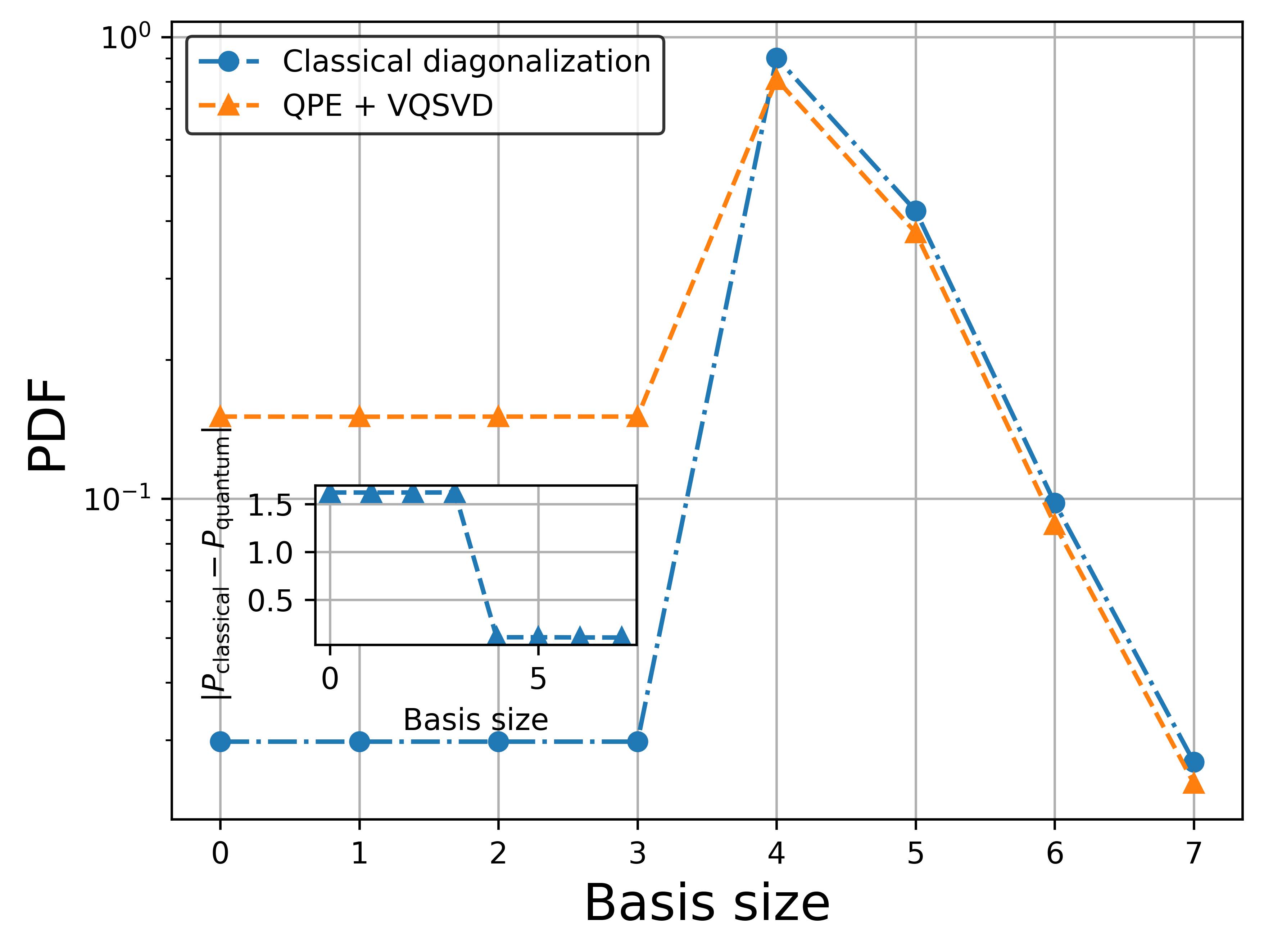} }}%
    \subfloat[$~V = 10.5$.]
    {{\includegraphics[width=0.33\textwidth]{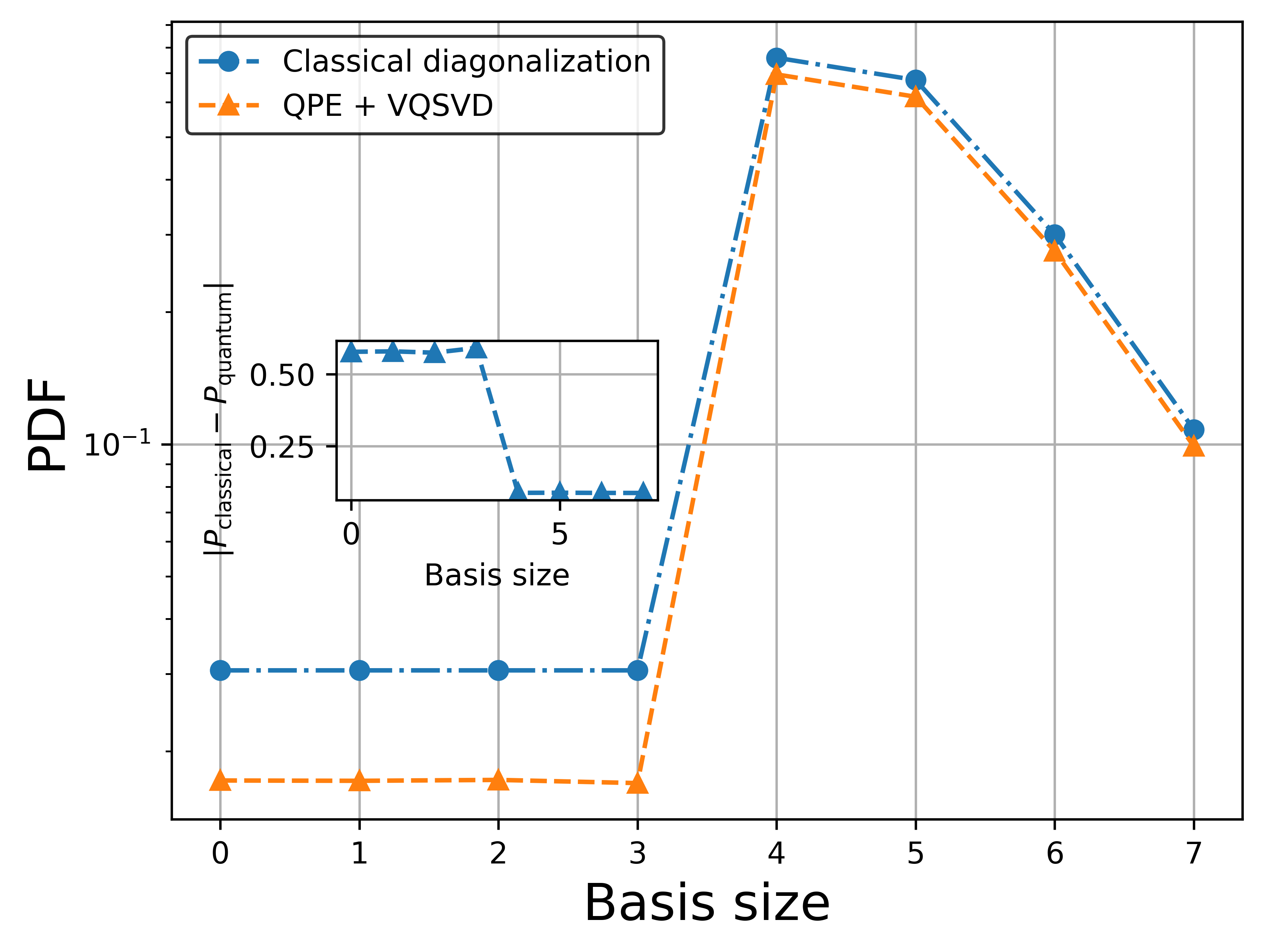} }}%
    \caption{Plots for the zeromode obtained via exact classical diagonalization and QPE + VQSVD on {\tt ibm\_brisbane} for different system volumes. We use three query qubits and seven precision qubits to construct the QPE circuit, respectively. The insets show the absolute error between the classical and quantum zeromodes for each basis size. Note that we plot both zeromodes on a log scale. Only the last four data points are relevant for our quantitative analysis.}
    \label{zeromodes}
\end{figure*}

We observe a reasonably good quantitative agreement between the classical and quantum results for each system volume considered [we obtain root-mean-square (RMS) deviation errors between the exact (classical diagonalization) and quantum zeromodes on the order of $3-5 \%$ (see the Supplementary Materials for more details)]. As mentioned previously, we employed only half of the basis set to compute the lowest eigenvalue of the stochastic Schl\"ogl matrix and its corresponding zeromode. Therefore, the first four data points in Fig. \ref{zeromodes} are irrelevant for our analysis. 

% We defer plots reporting a comparison of the RMS deviation between the classical and quantum zeromodes for different system volumes (across different DD sequences) to the Supplementary Materials. 

To compute the RMS deviation, we use
\begin{equation}
    \label{RMS deviation}
    \text{RMSD} = \sqrt{\frac{\displaystyle\sum_{i = 1}^{m} \left(y_i^{\text{exact}} - y_i^{\text{quantum}}\right)^2}{m}}, 
\end{equation}
where $y^{\text{exact}}$ ($y^{\text{quantum}}$) denotes the zeromode obtained via exact classical diagonalization (QPE + VQSVD) and $m$ denotes the number of data points over which we sample the distribution (four, in our case).

As a further check, we report results for the steady-state expectation value of the bistable Schl\"ogl operator $\hat{Q}_H$ [i.e., $\langle \hat{Q}_H \rangle = \bra{u_0} \hat{Q}_H \ket{u_0}$, where $\ket{u_0}$ denotes the non-equilibrium steady-state (zeromode)] for different system volumes in the Supplementary Materials. We expect $\langle \hat{Q}_H \rangle$ to identically vanish with respect to the non-equilibrium steady-state. Barring numerical floating point errors, we obtain values for $\langle \hat{Q}_H \rangle$ very close to zero (i.e., $\sim \mathcal{O}(10^{-4})$). This indicates that the results for the non-equilibrium steady-state we obtain via the implementation of QPE + VQSVD are consistent with the results we obtain via exact diagonalization for different system volumes. 

\section{Discussion}
\label{Discussion}

In this work, we applied different strategies to analyze the contributions of each term to the full Schl\"ogl operator with the goal of eliminating some of the less important terms. The errors encountered in implementing variational quantum algorithms such as VQD can be mitigated by considering operators with fewer Pauli terms that require fewer measurements. To understand the relevance of each term, we applied different sorting techniques and computed the expectation values of the resulting truncated operators. By employing different sorting techniques, we found that about half of the terms were not required to estimate an exact value of $\lambda_0$ for any Schl\"ogl operator constructed by Pauli decomposing Equation \ref{eq:Qspd}. Although we only present results for the bistable case in the main text, this was found to be surprisingly true for the monostable operator as well. No symmetry was used to directly reduce the number of gates in the circuit. Instead, the truncation of terms in an operator heavily depended on how the original operator was mapped to qubits. 
%For instance, using Equation \ref{eq:Qh} rather than Equation \ref{eq:Qspd} resulted in a completely different set of Pauli strings which could not be truncated in the same way.  
Additionally, no symmetries were identified in the operator for $\lambda_1$  that could reliably facilitate a cancellation or omission of terms. However, the magnitude sort yielded a reasonably good trade-off between the error and the number of terms to be dropped from the full operator. 

It is worth noting that the symmetries identified in the classical (Hermitian) operator also depend on the transformations used to construct a sparse representation of the operator and on the decomposition employed to express the classical operator as a linear combination of Pauli strings. For example, different sets of Pauli strings may arise with different symmetries, if any, when computing the eigenvalues of Equation \ref{eq:Qh} versus those of Equation \ref{eq:Qspd}. Therefore, investigating how a given operator is mapped to qubits is a crucial step that warrants further exploration for applications pertaining to the simulation of classical systems on near-term quantum hardware. 

The greatest computational gains in our simulations were seen using VQD with exact initial conditions, which converged in at least half the number of steps required to converge standard VQD with a naive initial state. This too is consistent with what is known about variational algorithms (both classical and quantum)---their accuracy and efficiency is highly dependent on the quality of that initial "guess." Our work exemplifies this fact and highlights the need to develop improved initial states for the quantum simulation of classical differential equations. The other requirement for quantum computing the eigenvalues for such a system is the existence of a reasonable Hermitian approximation of the system. The first non-zero eigenvalue is an important observable in a lot of physical applications, including stochastic kinetics. It is also very difficult to compute classically for realistic system sizes.\cite{bernstein_supersymmetry_1984, kalmykov_evaluation_2000} Developing variational quantum algorithms capable of extracting the lowest non-zero eigenvalue of a physical system will prove to be a fruitful venture.

We implemented QPE + VQSVD on the {\tt ibm\_brisbane} QPU to obtain an estimate for the zeromode of the Schl\"ogl operator for different system volumes and observed RMS deviation errors on the order of $3 - 5\%$. We note, however, that the basis size we used is insufficient to reproduce the bistability dynamics of the Schl\"ogl model. To approach the region of bistability for any given system volume, the use of a much larger basis size is warranted on both classical and quantum computers. Owing to resource constraints on current-NISQ hardware, we argue that the use of large basis sizes (i.e., the simulation of higher-dimensional stochastic Schl\"ogl matrices) on presently available quantum machines remains practically infeasible. It also remains unclear if the variational quantum algorithms employed in this paper (i.e., VQD and VQSVD) promise a quantum advantage. We hope to address this question via a detailed analysis of the scaling of classical versus quantum implementations in a future work. 

For the efficient simulation of higher-dimensional stochastic Schl\"ogl matrices on near-term quantum hardware, dimensionality reduction techniques may prove to be useful. We note that a systematic analysis of the number of independent degrees of freedom that VQSVD can efficiently handle (and by extension, the dimension of matrices that it can find the SVD of) is warranted. We suspect that an additional optimization of the existing VQSVD algorithm may be required to enable large-scale simulations.

\section{Conclusions and Outlook}
\label{Conclusions and outlook}
In this work, we have demonstrated how an important stochastic differential equation, the Chemical Master Equation, can be solved on quantum hardware using a variety of quantum algorithms suitable for both near-term and future fault-tolerant quantum hardware. In particular, we developed and applied quantum algorithms to compute the first two eigenvalues and the non-equilibrium steady state of the Schl\"ogl model, a paradigmatic example of a trimolecular reaction network known to exhibit multiple equilibrium states, using the VQD, QPE, and VQSVD algorithms, respectively. To do so, we demonstrated how the Schl\"ogl model's stochastic $Q$ matrix can be cast into a Hermitian form amenable to variational quantum eigensolvers and then employed classical modeling to benchmark which portions of the original non-Hermitian probability distributions can be captured in this way. The model's eigenvalues were subsequently calculated via noiseless and noisy quantum simulations for 2-, 3-, and 4-qubit Schl\"ogl operators. The quantum computed non-equilibrium steady-state agrees within a few percent of the non-equilibrium steady-state obtained via exact diagonalization for all system volumes considered. 

In our eigenvalue simulations, various sorting procedures were employed to analyze the contribution of each Pauli string to the expectation value of the full Schl\"ogl operator. For $\lambda_0$, we find that the operator can be truncated by at least half without sacrificing accuracy. While no such symmetry was identified for $\lambda_1$, sorting the Pauli strings by the magnitude of their coefficients yielded the lowest percentage errors: 0\% for the 2-qubit case with a 7/10 truncation ratio, and 1.29\% for the 3-qubit operator with a 24/28 truncation ratio. Notably, we numerically identify an exact form of the eigenvector corresponding to the lowest non-zero eigenvalue of the Hermitian Schl\"ogl operator. We use this exact eigenstate as the initial state in our implementation of a modified version of the VQD algorithm, VQD-{\tt exact0}. Using this exact initialization of the lowest state, we observe a significant reduction in the number of VQD iterations. Using 2-, 3-, and 4-qubit Schl\"ogl operators, we consistently observe at least a 50\% reduction in the number of classical optimization iterations. 

One of the outstanding challenges for variational quantum algorithms is the accumulation of errors due to the multiple measurements that need to be performed as the classical optimization loop progresses.\cite{fedorov_vqe_2022} This means that larger circuits with more parameters will incur more error than smaller circuits. Additionally, larger circuits demand a much more expressive ans\"atz with more parameters and more qubits. These facts are consistent with our results, as we observed the largest errors in the 4-qubit case. 

Reducing the number of measurements, the depth of the quantum circuit, and number of optimization parameters while maintaining accuracy are crucial for scaling variational quantum algorithms to realistic system sizes beyond simplistic toy models. Some classical systems possess operators with symmetries that can be exploited by quantum algorithms, as demonstrated in this paper for the Schl\"ogl model. In our work, we furthermore illustrate the importance of utilizing larger basis set sizes (i.e., higher-dimensional stochastic Schl\"ogl matrices) to reproduce the bistable dynamics of the (bistable) Schl\"ogl model. A smaller basis set is by no means sufficient to reproduce the bistable nature of the non-equilibrium steady-state, as we note in Secs. \ref{Zeromode} and \ref{Discussion}. To achieve this on both classical and quantum hardware, the use of a much larger basis set is warranted (see Sec. \ref{Discussion}). Thus, the development of variational quantum algorithms that can efficiently enable the simulation of such higher-dimensional matrices (and related system representations) is in order. We hope that our work spearheads further research in this direction.  

There is significant promise in advancing the application of quantum algorithms to classical problems, although this avenue remains underexplored. The zeromode approach allows one to directly target the non-equilibrium steady-state of a classical dynamical system (both deterministic and stochastic in nature). This enables the efficient estimation of relevant observables of interest in the non-equilibrium steady-state. In the context of this approach, we envision a scenario in which a smaller basis version of a classical problem may be analyzed using conventional classical algorithms to understand the form of the Hermitian zeromode of the classical dynamical system in question and the symmetries present therein. This information can subsequently be used to inform the initialization of the initial ``guess" (i.e., the initial quantum state) to be fed to a variational quantum algorithm to simulate the classical dynamical system on near-term quantum hardware using a larger basis set. 

We emphasize that developing better workflows (and refining existing ones) for solving non-Schr\"odinger type PDEs such as the CME is in great need. Enhanced initialization schemes are crucial to ensure the convergence and numerical stability of large-scale simulations of non-Schr\"odinger-type PDEs. Moreover, the development of improved variational quantum algorithms is essential for leveraging quantum computing capabilities to tackle non-Schr\"odinger-type PDEs efficiently. The pursuit of more exact theoretical and numerical approaches is also essential for achieving higher precision in modeling physical phenomena on near-term quantum hardware accurately. 

Our work opens new avenues for using current-NISQ devices as an alternative to classical computation to solve exponentially challenging stochastic chemical kinetics problems. We demonstrate how quantum algorithms can be extended to the simulation of stochastic chemical kinetics, thereby paving the way for new research directions and methodologies in both quantum computing and computational chemistry. It remains to be seen if a quantum advantage can be achieved in modeling stochastic chemical reactions and/or networks on near-term quantum hardware, and we hope that our work inspires further research in this direction. 

\section*{Acknowledgments}
\label{Ack}
T.K. (modeling, analysis, and manuscript preparation), Y.M.L. (modeling, analysis, and manuscript preparation), J.B.M. (concept, mentoring), and B.M.R. (concept, mentoring, and manuscript preparation) acknowledge support from the Brown University Office of the Vice-President for Research for seed funding for this project. We acknowledge D. Wei for providing computer code (in particular, the QPE subroutines) that was utilized in our analysis. We furthermore acknowledge the use of the IBMQ Experience for this work. The views expressed here are those of the authors and do not reflect the official policy and/or position of IBM or the IBMQ team.

\section*{Data and code availability}
Our \href{https://github.com/YashLokare02/Schlogl-model}{Schl\"ogl model} GitHub repository contains code to implement the classical and quantum subroutines. Additional data pertaining to the VQD ans\"atz and classical optimizer analysis (for VQD) can also be found in the aforementioned repository. 

\section*{Supplementary Materials}
We tabulate all of the data plotted and provide more detailed explanations of our parameter choices in our Supplementary Materials. 
 
\section*{References}
\bibliography{references}
% % %\bibliographystyle{unsrt}
\bibliographystyle{unsrtnat}

\end{document}

\onecolumn
\title{Supplementary Materials --- Modeling Stochastic Chemical Kinetics on Quantum Computers}
\date{}
\maketitle

\par\noindent\rule{\textwidth}{0.4pt}

\section{Qiskit Software Versions}

\subsection{VQD Implementation}
The Qiskit software versions used to implement VQD on local simulators are as reported below:
\begin{table}[H]
\centering
\begin{tabular}{|c|c|}
\hline
% \multicolumn{2}{|c|}{\textbf{Qiskit}} \\
% \hline
\multicolumn{1}{|c|}{\textbf{Qiskit software}} & \multicolumn{1}{|c|}{\textbf{Version}} \\
\hline
{\tt qiskit} & 0.44.1 \\
\hline
{\tt qiskit-terra} & 0.25.1 \\
\hline
{\tt python} & 3.10.12 \\
\hline
{\tt qiskit-algorithms} & 0.1.0 \\
\hline
\end{tabular}
\end{table}
We used the {\tt poetry Python} version package management system to manage all package dependencies. The {\tt .toml} file can be found on the \href{https://github.com/YashLokare02/Schlogl-model}{GitHub repository} along with the computer code generated to run numerical simulations and/or experiments.

\subsection{QPE + VQSVD Implementation}
The Qiskit software versions used to implement the QPE and VQSVD subroutines on {\tt ibm\_brisbane} are as reported below: 

\begin{table}[H]
\centering
\begin{tabular}{|c|c|}
\hline
% \multicolumn{2}{|c|}{\textbf{Qiskit}} \\
% \hline
\multicolumn{1}{|c|}{\textbf{Qiskit software}} & \multicolumn{1}{|c|}{\textbf{Version}} \\
\hline
{\tt qiskit-terra} & 0.24.0 \\
\hline
{\tt qiskit-aer} & 0.12.0 \\
\hline
{\tt qiskit-ibmq-provider} & 0.20.2 \\
\hline
{\tt qiskit} & 0.43.0 \\
\hline
\end{tabular}
\end{table}

\section{Non-Equilibrium Steady State (Stochastic Bistable Schl\"ogl Matrix)}
\subsection{$V = 1.1$}
\subsubsection{QPE + VQSVD results --- {\tt ibm\_brisbane}}
\begin{itemize}
    \item Number of experimental shots: $5 \times 10^5$; number of precision qubits: $7$; number of query qubits: $3$.  
    \item \textbf{Parameters set for training the quantum neural networks}: 
    \begin{enumerate}
    \item Classical optimizer: Adam optimizer
    \item Number of classical optimization iterations: 200
    \item Learning rate: 0.02
    \item Circuit depth: 55
    \item Weights for VQSVD: [24, 21, 18, 15, 12, 9, 6, 3]
    \end{enumerate}
\end{itemize} 

\subsubsection{Numerical results}
\begin{itemize}
    \item \textbf{RMS deviation (classical and quantum zeromodes)}: 3.375\%.
    \item $\langle \hat{Q}_H \rangle_{\ket{\psi_0}}$: $-4.54 \times 10^{-4}$ (here, $\ket{\psi_0}$ denotes the non-equilibrium steady-state).
\end{itemize}

\subsubsection{Minimum eigenvalue of the Schl\"ogl operator matrix extracted using QPE}

\begin{table}[H]
\centering
\caption{Minimum eigenvalue of the Schl\"ogl operator matrix extracted using QPE.}
\begin{tabular}{|c|c|}
\hline
\multicolumn{2}{|c|}{$\lambda_{\text{min}}$} \\
\hline 
\multicolumn{1}{|c|}{$\lambda_{\text{Schl\"ogl}}$} & \multicolumn{1}{|c|}{$\lambda_{\text{unitary}}$} \\
\hline 
0.05 & 0.999 + 0.04j \\
\hline 
\end{tabular}
\end{table}

% \begin{table}[H]
% \centering
%     \caption{RMS deviation between the classical and quantum zeromodes for $V = 1.1$ (simulation(s) run on {\tt ibm\_brisbane})}
%     \begin{tabular}{|l|c|}
%         \hline
%         \textbf{DD sequences} & \textbf{RMS deviation (\%)} \\
%         \hline
%         No DD & $3.375$ \\
%         \hline
%         Hahn-X & $1.665$ \\
%         \hline
%         Hahn-Y & $1.665$ \\
%         \hline
%         CP & $3.375$ \\
%         \hline
%         CPMG & $3.375$ \\
%         \hline
%         XYXY & $3.375$ \\
%         \hline
%         YZYZ & $3.375$ \\
%         \hline
%         XZXZ & $3.375$ \\
%         \hline
%         CDD (order 2) & $3.375$ \\
%         \hline
%         XY8 & $3.375$ \\
%         \hline
%         XY16 & $5.923$ \\
%         \hline
%         Uhrig-X & $3.375$ \\
%         \hline
%         Uhrig-Y & $3.375$ \\
%         \hline
%         KDD & $3.375$ \\
%         \hline
%     \end{tabular}
% \end{table}

% \begin{figure}[htbp]
% \includegraphics[width = \linewidth]{images/RMS_1.1.png}% Here is how to import EPS art
% \caption{(Color outline) Comparison of the RMS deviation between the classical and quantum zeromodes for $V = 1.1$ with DD (we use three query qubits and seven precision qubits to construct the QPE circuit, respectively).}
% \label{fig:rms1}
% \end{figure}

\subsection{$V = 5.5$}
\subsubsection{QPE + VQSVD results --- {\tt ibm\_brisbane}}
\begin{itemize}
    \item Number of experimental shots: $5 \times 10^5$; number of precision qubits: $7$; number of query qubits: $3$.  
    \item \textbf{Parameters set for training the quantum neural networks}: 
    \begin{enumerate}
    \item Classical optimizer: Adam optimizer
    \item Number of classical optimization iterations: 200
    \item Learning rate: 0.02
    \item Circuit depth: 55
    \item Weights for VQSVD: [24, 21, 18, 15, 12, 9, 6, 3]
    \end{enumerate}
\end{itemize} 

\subsubsection{Numerical results}
\begin{itemize}
    \item \textbf{RMS deviation (classical and quantum zeromodes)}: 5.141\%.
    \item $\langle \hat{Q}_H \rangle_{\ket{\psi_0}}$: $1.47 \times 10^{-4}$ (here, $\ket{\psi_0}$ denotes the non-equilibrium steady-state).
\end{itemize}

\subsubsection{Minimum  eigenvalue of the Schl\"ogl operator matrix extracted using QPE}

\begin{table}[H]
\centering
\caption{Minimum eigenvalue of the Schl\"ogl operator matrix extracted using QPE.}
\begin{tabular}{|c|c|}
\hline
\multicolumn{2}{|c|}{$\lambda_{\text{min}}$} \\
\hline 
\multicolumn{1}{|c|}{$\lambda_{\text{Schl\"ogl}}$} & \multicolumn{1}{|c|}{$\lambda_{\text{unitary}}$} \\
\hline 
0.04 & 0.999 + 0.05j \\
\hline 
\end{tabular}
\end{table}

% \subsubsection{RMS deviation between the classical and quantum zeromodes (different DD sequences)}

% \begin{table}[H]
% \centering
%     \caption{RMS deviation between the classical and quantum zeromodes for $V = 5.5$ (simulation(s) run on {\tt ibm\_brisbane})}
%     \begin{tabular}{|l|c|}
%         \hline
%         \textbf{DD sequences} & \textbf{RMS deviation (\%)} \\
%         \hline
%         No DD & $5.141$ \\
%         \hline
%         Hahn-X & $5.719$ \\
%         \hline
%         Hahn-Y & $5.719$ \\
%         \hline
%         CP & $4.915$ \\
%         \hline
%         CPMG & $4.915$ \\
%         \hline
%         XYXY & $4.672$ \\
%         \hline
%         YZYZ & $5.216$ \\
%         \hline
%         XZXZ & $5.216$ \\
%         \hline
%         CDD (order 2) & $5.216$ \\
%         \hline
%         XY8 & $5.216$ \\
%         \hline
%         XY16 & $4.856$ \\
%         \hline
%         Uhrig-X & $4.672$ \\
%         \hline
%         Uhrig-Y & $5.377$ \\
%         \hline
%         KDD & $5.421$ \\
%         \hline
%     \end{tabular}
% \end{table}

% \begin{figure}[htbp]
% \includegraphics[width = \linewidth]{images/RMS_5.5.png}% Here is how to import EPS art
% \caption{(Color outline) Comparison of the RMS deviation between the classical and quantum zeromodes for $V = 5.5$ with DD (we use three query qubits and seven precision qubits to construct the QPE circuit, respectively).}
% \label{fig:rms2}
% \end{figure}

\subsection{$V = 10.5$}
\subsubsection{QPE + VQSVD results --- {\tt ibm\_brisbane}}
\begin{itemize}
    \item Number of experimental shots: $5 \times 10^5$; number of precision qubits: $7$; number of query qubits: $3$.  
    \item \textbf{Parameters set for training the quantum neural networks}: 
    \begin{enumerate}
    \item Classical optimizer: Adam optimizer
    \item Number of classical optimization iterations: 200
    \item Learning rate: 0.02
    \item Circuit depth: 55
    \item Weights for VQSVD: [24, 21, 18, 15, 12, 9, 6, 3]
    \end{enumerate}
\end{itemize} 

\subsubsection{Numerical results}
\begin{itemize}
    \item \textbf{RMS deviation (classical and quantum zeromodes)}: 4.454\%.
    \item $\langle \hat{Q}_H \rangle_{\ket{\psi_0}}$: $3.24 \times 10^{-4}$ (here, $\ket{\psi_0}$ denotes the non-equilibrium steady-state).
\end{itemize}

\subsubsection{Minimum eigenvalue of the Schl\"ogl operator matrix extracted using QPE}

\begin{table}[H]
\centering
\caption{Minimum eigenvalue of the Schl\"ogl operator matrix extracted using QPE.}
\begin{tabular}{|c|c|}
\hline
\multicolumn{2}{|c|}{$\lambda_{\text{min}}$} \\
\hline 
\multicolumn{1}{|c|}{$\lambda_{\text{Schl\"ogl}}$} & \multicolumn{1}{|c|}{$\lambda_{\text{unitary}}$} \\
\hline 
0.03 & 0.999 + 0.03j \\
\hline 
\end{tabular}
\end{table}

\pagebreak

\subsection{Loss Curves for VQD with Exact Initial State (VQD-{\tt{exact0}})}

\begin{figure*}[h]
    \centering
    {{\includegraphics[width=0.31\textwidth]{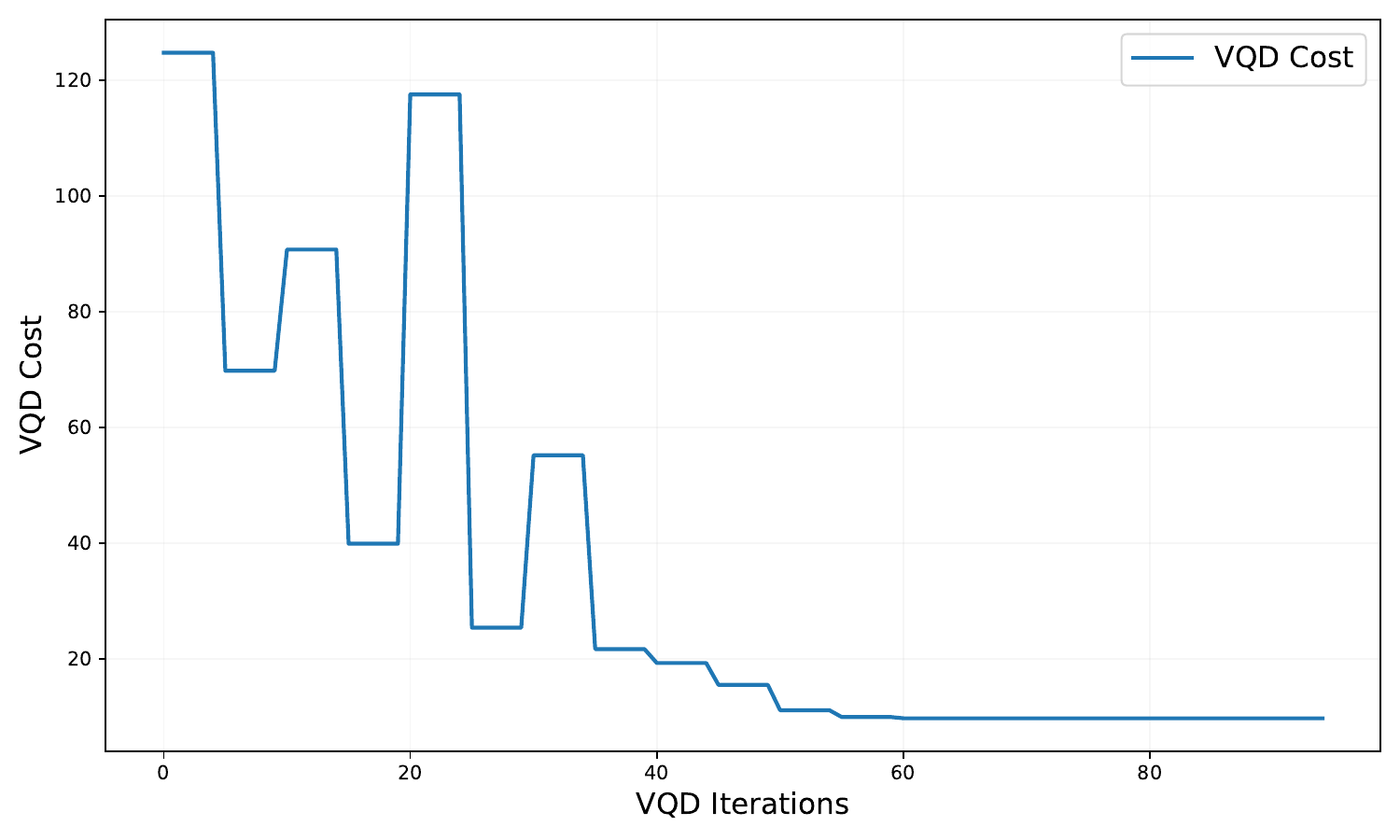} }}%
    {{\includegraphics[width=0.31\textwidth]{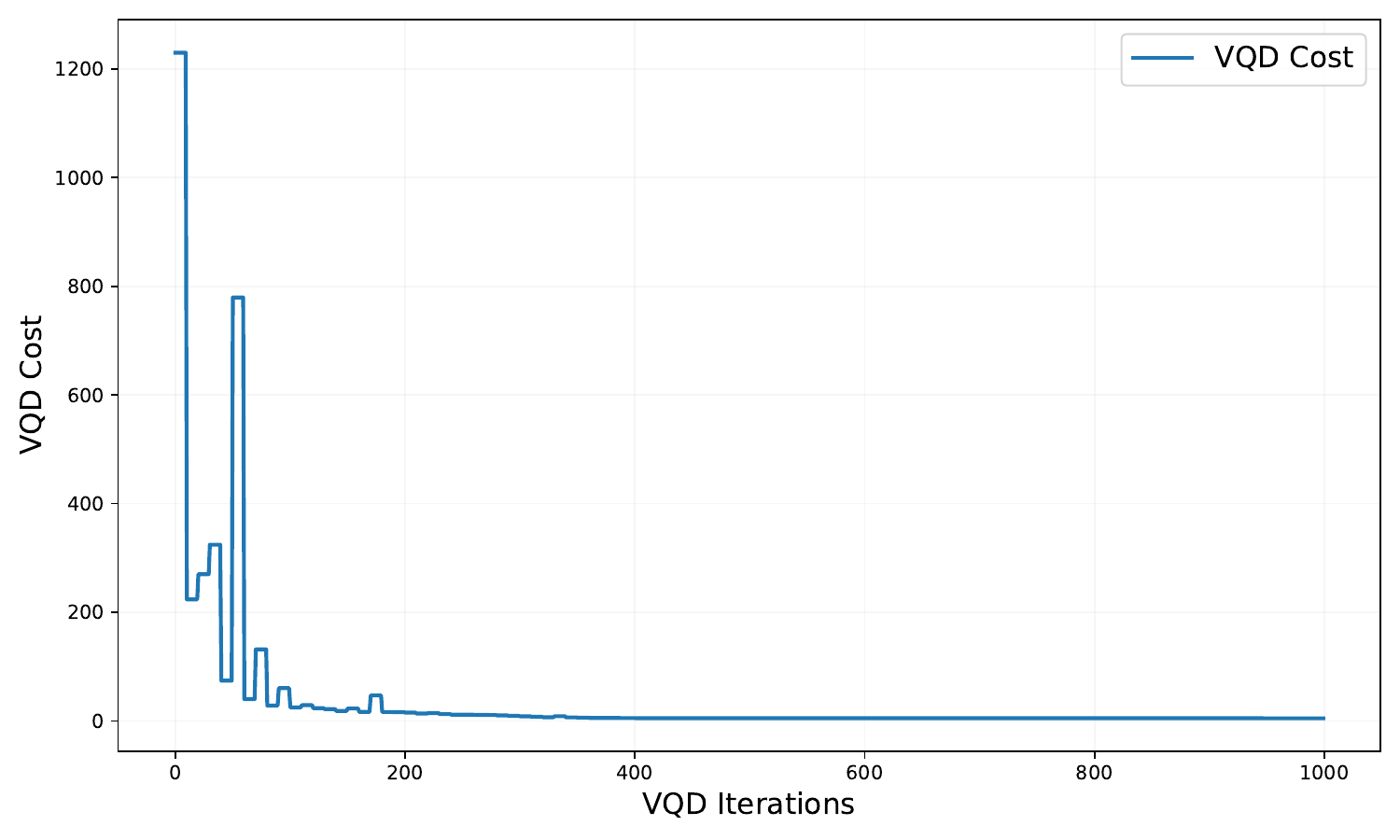} }}%
    {{\includegraphics[width=0.31\textwidth]{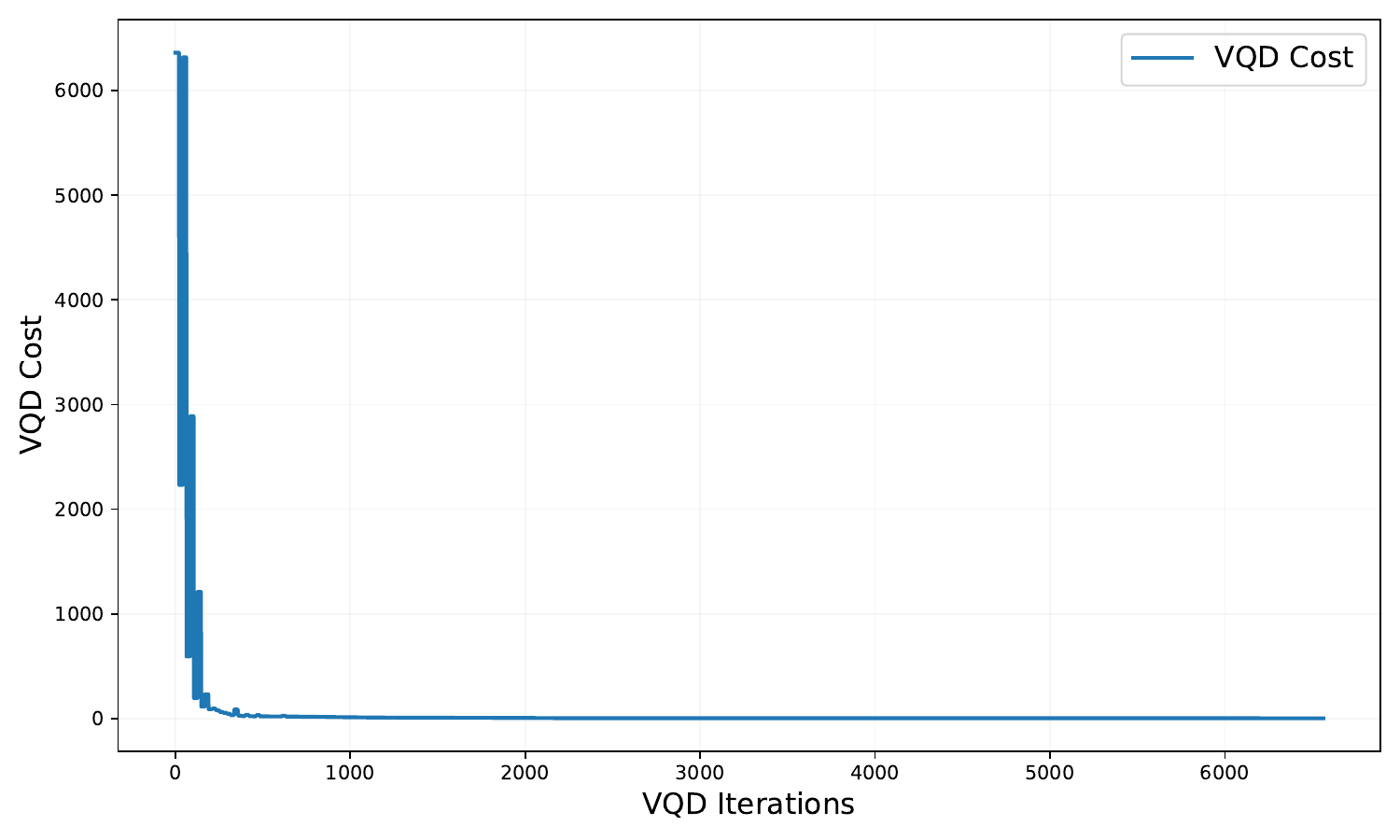} }}%
    \caption{Minimization of the VQD cost function for different qubit bistable operators. (a) VQD-{\tt{exact0}} cost with 2-qubit bistable operator. (b) VQD-{\tt{exact0}} cost with 3-qubit bistable operator. (c) VQD-{\tt{exact0}} with 4-qubit bistable operator.}
    
    \label{vqd-cost}
\end{figure*}

%\pagebreak

\subsection{Loss Curves (VQSVD)}

\begin{figure*}[h]
    \centering
    \subfloat[$V = 1.1$.]
    {{\includegraphics[width=0.33\textwidth]{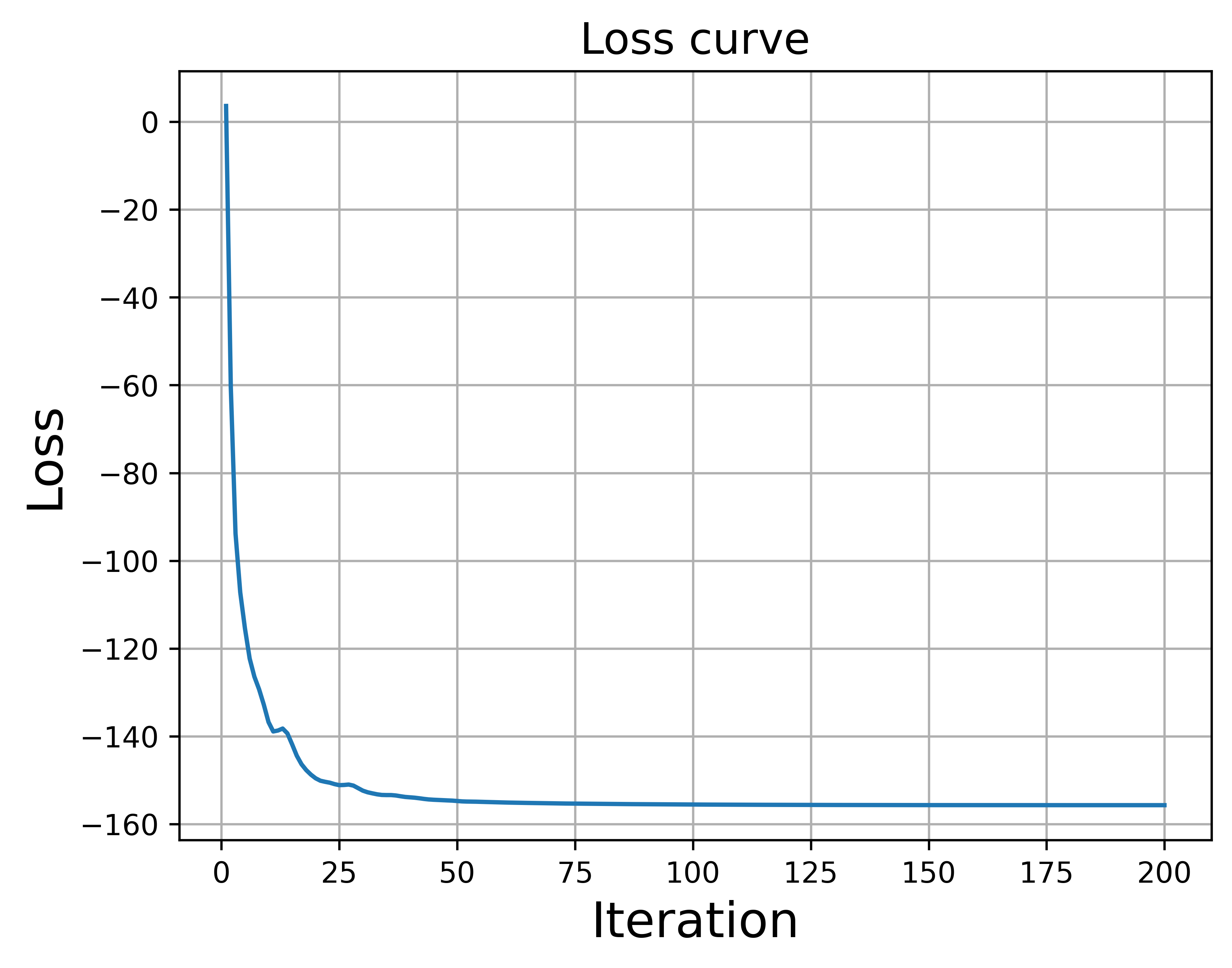} }}%
    \subfloat[$V = 5.5$.]
    {{\includegraphics[width=0.33\textwidth]{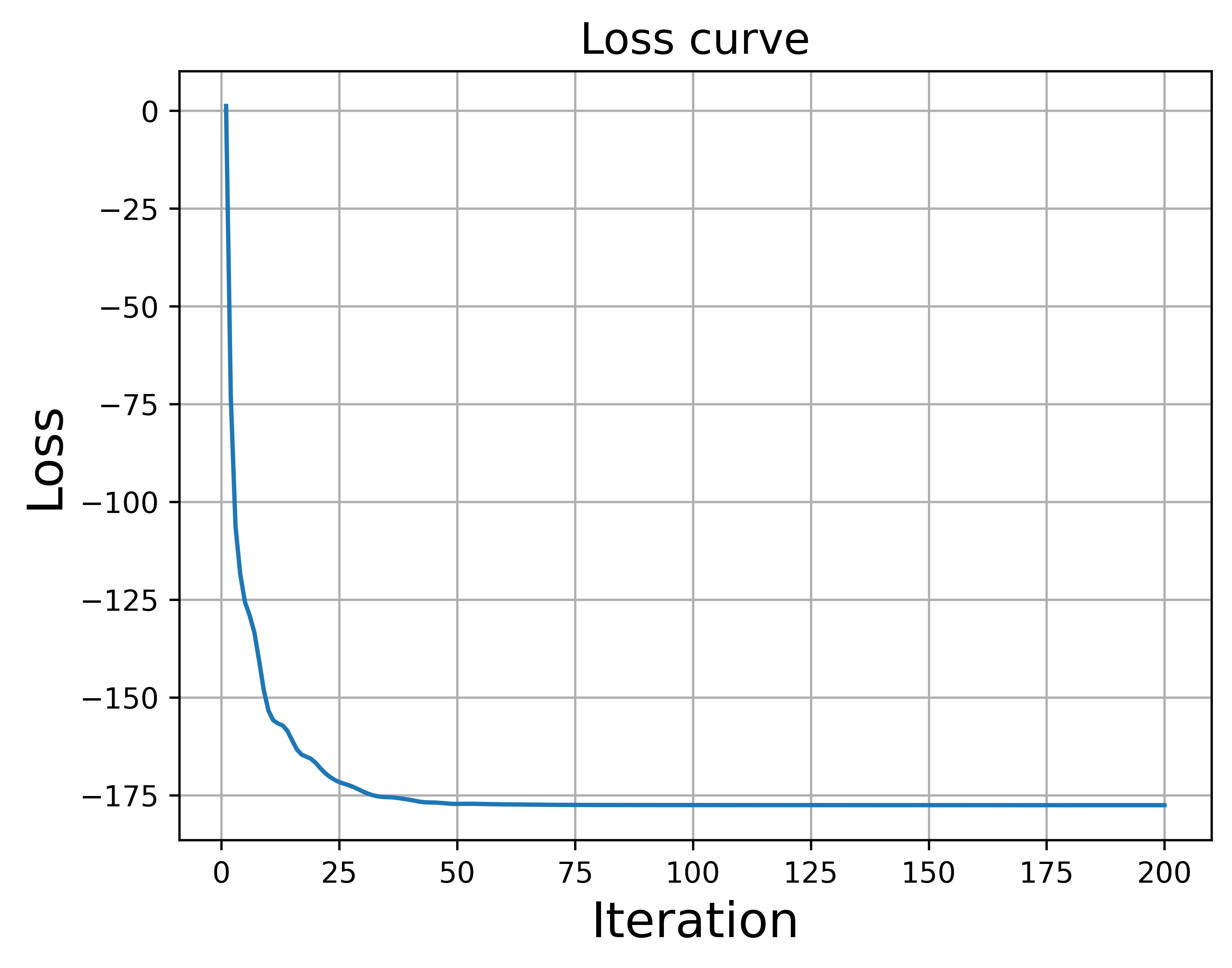} }}%
    \subfloat[$V = 10.5$.]
    {{\includegraphics[width=0.33\textwidth]{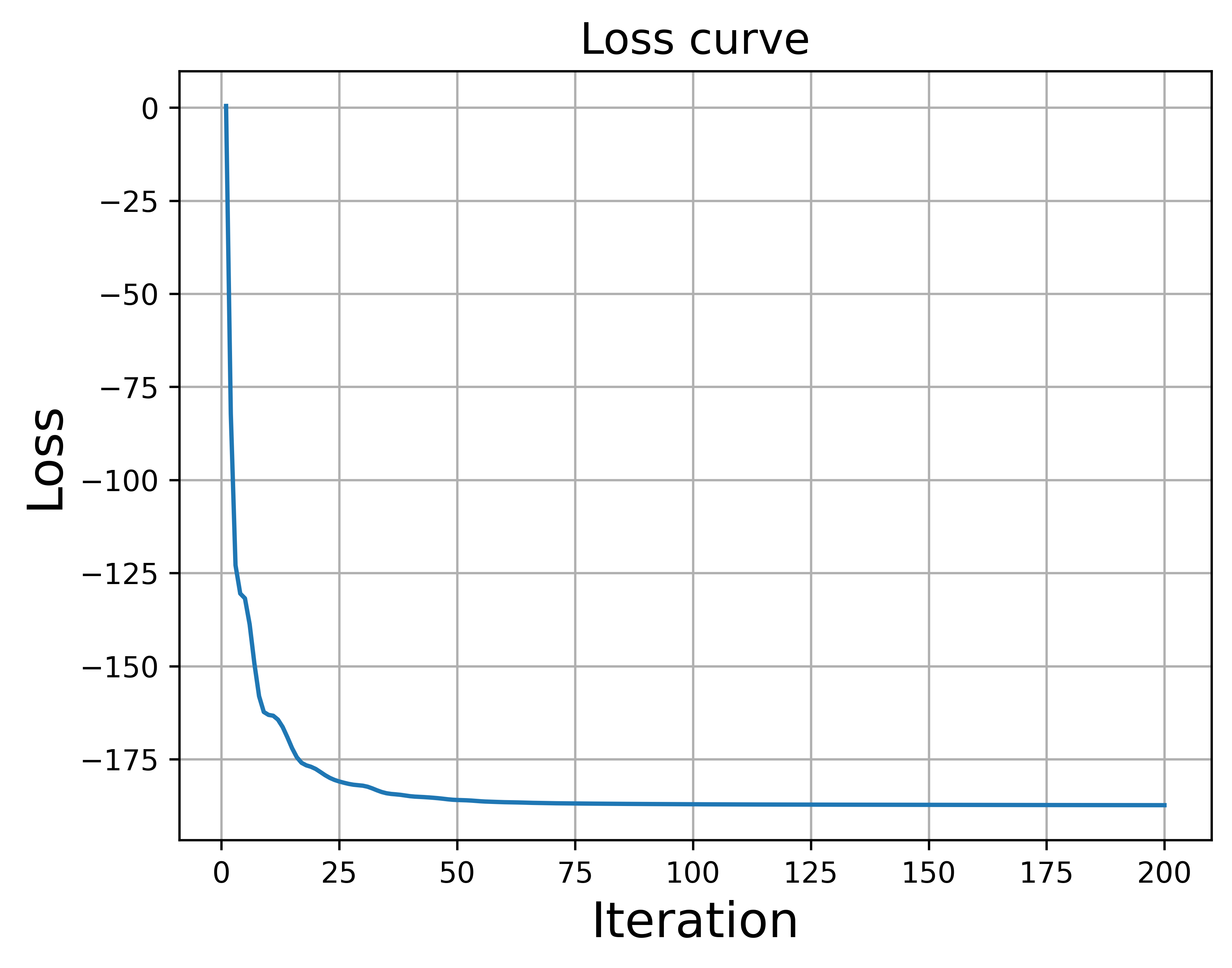} }}%
    \caption{Minimization of the VQSVD cost function for different system volumes (simulations run on {\tt ibm\_brisbane}).}
    \label{VQSVD cost}
\end{figure*}

\pagebreak

\subsection{Percentage Error for Eigenvalues Table}

%% Table showing percentage errors in lambda1

\setlength{\tabcolsep}{13pt}
\begin{table}[H]
  \centering
  \begin{tabular}{ccccc}
    \toprule
    Operator & Positive-First Sort & Default Sort & Magnitude Sort & Optimized Sort \\
    \midrule
    & & & & \\  
    \multirow{2}{*}{2-qubit $\lambda_0$} & $0.00$ (10) & $0.00$ (10) & $0.00$ (10) & $-$ \\
     & $0.79$ (7) & $0.00$ (6) & $0.00$ (8) & $0.00$ (4)\\
    & & & & \\  
    \multirow{2}{*}{2-qubit $\lambda_1$} & $0.00$ (10) & $0.00$ (10) & $0.00$ (10) & $-$\\
     & $-$ & $-$ & $0.00$ (7) & $-$\\
    
     \hline
     & & & & \\  
    \multirow{2}{*}{3-qubit $\lambda_0$} & $0.00$ (28) & $0.00$ (14) & $0.00$ (28) & $0.00$ (6) \\
     & $0.55$ (4) & $-$ & $-$ & $-$\\
     & & & & \\
    \multirow{2}{*}{3-qubit $\lambda_1$} & $0.00$ (28) & $0.00$ (28) & $0.00$ (28) & $-$ \\
     & $39.06$ (20) & $27.77$ (23) & $1.29$  (24) & $-$ \\ 
     & & & & \\  
   % \hline
   %  & & & & \\  
   %  \multirow{1}{*}{4-qubit $\lambda_0$} & $-$ & $0.00$ (36) & $-$ & $0.00$ (8)\\
   %  \multirow{1}{*}{4-qubit $\lambda_1$} & $-$ & $1.46$ (72) & $-$ & $-$ \\
   %  & & & & \\  
    \bottomrule
  \end{tabular}
  \caption{Percentage error in $\lambda_0$ and $\lambda_1$ for the 2- and 3-qubit operators. For each qubit case, the first row represents results from the full operator while the second row denotes results from the respective truncation method. ``$-$" means that the error in the respective truncation method was significant, i.e., larger than 100\%. The digits in parentheses denote the number of Pauli terms used in the operator. The best absolute error and percentage error for $\lambda_0$ and $\lambda_1$ is given for each truncation method of the full operator. The error plots supporting this table are shown in Figure 5 in the main text.}
  \label{tab:error}
\end{table}